\documentclass[aps,pra,twocolumn,twoside,superscriptaddress,
               nofootinbib]{revtex4} \hbadness2000

\usepackage{graphicx}
\usepackage{amsmath}
\usepackage{amssymb}

\newcommand{\nc}{\newcommand}
\nc{\rnc}{\renewcommand}

\def\os2{\frac{1}{\sqrt{2}}}
\nc{\smfrac}[2]{\mbox{$\frac{#1}{#2}$}}
\def\sos2{\smfrac{1}{\sqrt{2}}}
\nc{\Tr}{\operatorname{Tr}}
\nc{\ox}{\otimes}
\nc{\dg}{\dagger}
\def\C{{\cal C}}
\def\t{\theta}

\def\ot{\otimes}
\nc{\eq}[1]{(\ref{eq:#1})}
\nc{\eqs}[2]{(\ref{eq:#1}) and (\ref{eq:#2})}
\nc{\eqm}[2]{(\ref{eq:#1})--(\ref{eq:#2})}
\rnc{\sec}[1]{Sec.~\ref{sec:#1}}
\nc{\be}{\begin{equation}}
\nc{\ee}{{\end{equation}}}
\nc{\bea}{\begin{eqnarray}}
\nc{\eea}{\end{eqnarray}}
\nc{\<}{\langle}
\rnc{\>}{\rangle}

\def\lpm{ \left(\rule{0pt}{2.1ex}\right. \!}
\def\rpm{ \!\left.\rule{0pt}{2.1ex}\right) }

\def\non{\nonumber}
\nc\ms{\hspace*{-0.1ex}}

\def\fb{\framebox}
\def\mb{\makebox}

\nc{\setl}[1]{\setlength{\unitlength}{#1}  \centering}
\def\bp{\begin{array}{c} \begin{picture}}
\def\ep{\end{picture}\end{array}}

\begin{document}

\title{Unified derivations of measurement-based schemes for quantum
computation}

\author{Andrew M. Childs}
\email[]{amchilds@mit.edu}
\affiliation{Center for Theoretical Physics,
             Massachusetts Institute of Technology,
             Cambridge, MA 02139, USA}
\author{Debbie W. Leung}
\email[]{wcleung@caltech.edu}
\affiliation{Institute for Quantum Information,
             California Institute of Technology,
             Pasadena, CA 91125, USA}
\author{Michael A. Nielsen} 
\email[]{nielsen@physics.uq.edu.au} 
\affiliation{School of Physical Sciences and
             School of Information Technology \& Electrical Engineering,
             The University of Queensland,
             QLD 4072, Australia}

\date[]{26 June 2004}

\begin{abstract}
  We present unified, systematic derivations of schemes in the two
  known measurement-based models of quantum computation.
  The first model (introduced by Raussendorf and Briegel [Phys.\
  Rev.\ Lett.\ \textbf{86}, 5188 (2001)]) uses a fixed entangled
  state, adaptive measurements on single qubits, and feedforward of
  the measurement results.
  The second model (proposed by Nielsen [Phys.\ Lett.\ A \textbf{308}, 96
  (2003)] and further simplified by Leung [Int.\ J.\ Quant.\ Inf.
  \textbf{2}, 33 (2004)]) uses adaptive two-qubit measurements
  that can be applied to arbitrary pairs of qubits, and feedforward of
  the measurement results.
  The underlying principle of our derivations is a variant of
  teleportation introduced by Zhou, Leung, and Chuang 
  [Phys.\ Rev.\ A \textbf{62}, 052316 (2000)].
  Our derivations unify these two measurement-based models of quantum
  computation and provide significantly simpler schemes.
\end{abstract}
\maketitle

\parskip 1ex

%%%%%%%%%%%%%%%%%%%%%%%%%%%%%%%%%%%%%%%%%%%%%%%%%%%%%%%%%%%%%%%%%%%%%%%%%
\section{Introduction}

What physical resources are needed to simulate the evolution of an
arbitrary physical system?
In the context of information processing, the ability to perform {\em
universal} computation is equivalent to the ability to simulate an
arbitrary evolution---any computation is performed by evolving a
computing machine, and conversely, a universal computing machine can
be used to simulate the evolution of any system.

In the standard quantum circuit model of quantum computation
\cite{DiVincenzo95a,Preskill98bk,Nielsen00bk}, a quantum computation
involves initializing quantum systems---typically \emph{qubits} or
two-level quantum systems---that are then acted on by a sequence of
\emph{quantum gates}, followed by some measurements.
In this model, simple quantum gates (say, acting on one or two qubits
at a time) can be used to build up an \emph{arbitrary} unitary
transformation.
Nonunitary evolution such as dissipation can also be simulated in this
model by introducing and later discarding ancilla qubits.

Since measurement is generally irreversible, until recently the
conventional wisdom has held that the processing of quantum
information should be kept coherent and measurements should be delayed
until the final read-out of computation results.
A notable exception to this rule of thumb is quantum
teleportation~\cite{Bennett93}, in which a measurement by one party
determines the correction that a remote party should apply to recover
a quantum state.
Another notable exception is the use of syndrome measurements in
quantum error correction \cite{Sho95}.  Syndrome measurements reveal
the error that has occurred without measuring the encoded quantum
state, thereby preserving its coherence.
Indeed, there are many approaches to fault-tolerant quantum
computation in which measurements and simple quantum gates are used to
implement other quantum gates that are difficult to apply
directly~\cite{Shor96,Boykin99,Knill98,Gottesman99t,Zhou00,Knill01a,Lidar03}.

Raussendorf and Briegel~\cite{Raussendorf01a} overturned this
conventional wisdom, showing that it is possible to perform universal
quantum computation using a \emph{sequence of single-qubit
measurements alone}, acting on some fixed entangled state called a
{\em cluster state}~\cite{Raussendorf00}.
Once the cluster state is prepared, no further interactions are
required, and the only aspect of the computation that must remain
coherent is the storage of quantum information.
More precisely, \emph{any} quantum circuit up to depth $d$ and breadth
$b$ may be simulated using a single, fixed cluster state of $O(bd)$
qubits.  
Each simulation of a quantum gate is successful up to an additional
known Pauli error.
Since the act of measuring the cluster state is irreversible, this
model is referred to as the {\em one-way quantum computer} (1WQC)
model.

After the 1WQC was introduced, a very different measurement-based
scheme for quantum computation was introduced by one of
us~\cite{Nielsen01t}, following the line of thought developed
in~\cite{Nielsen97c}, \cite{Gottesman99t}, and \cite{Knill01a}.  
We will refer to this model as a \emph{teleportation-based model of
quantum computation} (TQC), since it is conceptually derived from
teleportation.
The TQC uses similar physical resources to the 1WQC: (multiple-qubit)
measurements, quantum memory, and feedforward.
The initial TQC scheme proposed in~\cite{Nielsen01t} uses four-qubit
measurements.  It also requires a nondeterministic number of steps to
perform each quantum gate.  
Simpler TQC schemes were later proposed
\cite{Fenner01,Leung01c,Leung03t}, with the simplest using only
two-qubit measurements and performing each gate deterministically (up
to a known Pauli error).

The TQC is easy to understand since it is similar to the standard 
model of quantum computation.
In comparison, the conceptual basis for the 1WQC is less clear.  The
prescriptions given in~\cite{Raussendorf01a,Raussendorf02a} for using
a 1WQC can be easily verified, but there is no clear underlying
principle.  This makes it nontrivial to modify or optimize the
existing 1WQC schemes.  It is also unclear what makes the cluster
state a good substrate for quantum computation, and more generally,
what makes a good or bad substrate.  Finally, the 1WQC formalism is
heavily based on the stabilizer language developed by
Gottesman~\cite{Gottesman96}.  Although this language is powerful, it
is also rather specialized, being limited primarily to the analysis of
situations in which operations from a special set---the Clifford
group---are being applied.  Furthermore, the connection between the
stabilizer language and the corresponding picture in terms of state
vectors is not always straightforward.

On the other hand, the 1WQC has important advantages over the TQC.
First, no quantum interactions are required after the initial
preparation of the cluster state.  Second, the cluster state is
independent of the computation to be performed, except for its breadth
and depth: arbitrary interactions can be extracted from the fixed
cluster state.  Third, there may be physical systems in which a
cluster state offers experimental advantages over more conventional
approaches~\cite{Raussendorf00,Nielsen04}.

Our initial goal in undertaking the research reported here was to
identify simple underlying principles for the 1WQC and to
systematically derive schemes similar to the proposed 1WQC.
We eventually found such a systematic derivation using teleportation
as an underlying principle \cite{ChildsPItalk03},
in accord with the conjecture that the 1WQC and the TQC are closely
related.
Our improved conceptual understanding of the 1WQC proved valuable, for
we subsequently found much simpler 1WQC-like
schemes~\cite{LeungIQItalk04} by choosing a simpler underlying
principle, known as ``one-bit teleportation'' \cite{Zhou00}.
Such simplification is reminiscent of the work in~\cite{Zhou00}, which
simplifies the systematic fault-tolerant gate construction proposed
in~\cite{Gottesman99t}.
We then realized that one-bit teleportation also simplifies schemes in
the TQC model~\cite{LeungERATOtalk04}.  

We have therefore unified the 1WQC and the TQC models and obtained
simplified measurement-based quantum computation schemes.  The 1WQC
schemes we derive combine the conceptual simplicity of the TQC with
the practical advantages of the 1WQC.  We have also identified one-bit
teleportation as a single principle underlying all existing approaches
to measurement-based quantum computation.

During the course of our investigation and preparation of the
manuscript, several related results have been reported.  
A different explanation of the 1WQC model in terms of valence bond
solids was reported by Verstraete and Cirac \cite{Verstraete03}.
Whereas our 1WQC-like schemes differ from the original schemes of
\cite{Raussendorf01a,Raussendorf02a}, an exact explanation of the
latter in terms of teleportation was given by Aliferis and one of us
\cite{Aliferis04}.
A partial explanation of the 1WQC model in terms of one-bit
teleportation was reported very recently by Jorrand and Perdrix
\cite{JP04}, while schemes similar to our simplified TQC schemes were
independently reported by Perdrix \cite{Perdrix04}.
Several results announced after our initial posting may also be of
interest.  These include a model of measurement-based universal
quantum Turing machines \cite{PJ04b}, further results on combining the
1WQC with linear optics \cite{Browne04} and a new fault-tolerance
study in the 1WQC \cite{ND04}.

The structure of the paper is as follows.  We begin in \sec{models} by
briefly reviewing the circuit model, introducing a notion of circuit
simulation with Pauli errors, and describing the TQC and 1WQC models
of quantum computation in more detail.
We emphasize some conceptual ideas that may be useful elsewhere.
\sec{1btsims} reviews one-bit teleportation and presents useful techniques
obtained from it.
The techniques are used to derive a simplified TQC scheme in \sec{newtqc}.  
\sec{1wqctg} explains how one-bit teleportation can be used to derive
a scheme that simulates arbitrary circuits using only an entangled
initial state, single-qubit measurements, and feedforward.  Here the
initial state depends on the circuit to be simulated.
\sec{universal} explains several techniques to remove the dependence
of the initial state on the circuit being simulated (except for its
breadth and depth).  
A short conclusion of the paper is given in \sec{conclusion}.
Our initial systematic derivation of 1WQC-like schemes based on
teleportation can be found in \cite{ChildsPItalk03}.

%%%%%%%%%%%%%%%%%%%%%%%%%%%%%%%%%%%%%%%%%%%%%%%%%%%%%%%%%%%%%%%%%%%%%
\section{The quantum circuit model, the TQC, and the 1WQC}
\label{sec:models}

In this section, we summarize the circuit model of quantum computation
as a way of introducing the notation used throughout the paper.  We
describe a notion of circuit simulation that is crucial to our
discussion.  Then, we outline the main features of the TQC and the
1WQC that motivate our derivation.  
Since we derive simplified TQC and 1WQC schemes, a full
summary of the existing schemes will be omitted.
Finally, we describe a diagrammatic representation of the 1WQC, 
which we call the {\em substrate representation}.

%%%%%%%%%%%%%%%%%%%%%%%%%%%%%%%%%%%%%%%%%%%%%%%%%%%%%%%%%%%%%%%%%%%%%%%%%
\subsection{The quantum circuit model}
\label{sec:circuitmodel}

Any unitary evolution can be built from simple quantum gates (say,
acting on one or two qubits at a time).  A circuit diagram represents
a sequence of unitary evolutions and measurements.  The input states
and measurement outcomes may be included.  In a circuit diagram, time
runs from left to right.  Each horizontal line represents quantum
information propagating forward in time, or equivalently, quantum
storage.  Often, each line represents one qubit of quantum
information.  A unitary gate is represented by a box on the line(s),
and a symbol for the gate is written inside the box.  Thus, the
circuit symbol for a single-qubit gate $U$ is given by 
\bea
\setl{0.6mm}
%\bp(30,20)
%\put(0,10){\line(1,0){10}}
%\put(10,5){\fb(10,10){$U$}}
%\put(20,10){\line(1,0){10}}
\bp(30,12)
\put(0,6){\line(1,0){10}}
\put(10,1){\fb(10,10){$U$}}
\put(20,6){\line(1,0){10}}
\ep
\label{eq:u}
\eea
In the basis $\{|0\>,|1\>\}$, the matrix representations of some
useful single-qubit gates are given by
\bea
        X_\t = e^{-i\t X} \,, ~
        Z_\t = e^{-i\t Z} \,, ~
        H = \smfrac{1}{\sqrt{2}} 
            \left( \begin{array}{cc} 1 & 1 \\ 1 & -1 \end{array} \right) ,\,
\eea
where $I,X,Y,Z$ are used to represent the Pauli operators
\bea
I 
= \left( \begin{array}{cc} 1 & 0 \\ 0 & 1 \end{array} \right) \,, &&  
X
= \left( \begin{array}{cc} 0 & 1 \\ 1 & 0 \end{array} \right) \,,
\non
\\
Y
= \left( \begin{array}{cc} 0 &-i \\ i & 0 \end{array} \right) \,, &&
Z
= \left( \begin{array}{cc} 1 & 0 \\ 0 &-1 \end{array} \right) \,.
\eea

The action of the Pauli operators on each qubit generates a group,
called the Pauli group.  The Clifford group consists of those unitary
operators that preserve the Pauli group by conjugation.
For example, 
\bea
	HXH = Z \,,~~HZH = X \,.
\label{eq:hadamard}
\eea
The only two-qubit gates we will use are within the Clifford
group---for example, the \emph{controlled-phase} and the
\emph{controlled-{\sc not}} gates.  They are denoted by $\Lambda(Z)$ and
$\Lambda(X)$ respectively, and their circuit symbols are given by
\bea
\setlength{\unitlength}{0.6mm}
\centering
\bp(40,30)
\put(-5,10){\mb(10,10){$\Lambda(Z)$:}}
\put(10,5){\line(1,0){20}}
\put(10,25){\line(1,0){20}}
\put(20,5){\line(0,1){20}}
\put(17,2){\line(1,1){6}}
\put(17,8){\line(1,-1){6}}
\put(17,22){\line(1,1){6}}
\put(17,28){\line(1,-1){6}}
\ep
\hspace*{5ex}
\bp(40,30)
\put(-5,10){\mb(10,10){$\Lambda(X)$:}}
\put(10,5){\line(1,0){20}}
\put(10,25){\line(1,0){20}}
\put(20,2){\line(0,1){23}}
\put(20,5){\circle{6}}
\ep
\label{eq:2qubitgates}
\eea
The target of $\Lambda(X)$ is taken to be the second qubit.  The
``upside down'' controlled-{\sc not} with the first qubit as the
target is denoted by $V(X)$.
In contrast, $\Lambda(Z)$ is symmetric between the two qubits, as is evident in
the notation of \eq{2qubitgates}.
In the basis $\{|00\>$, $|01\>$, $|10\>$, $|11\>\}$, the matrix
representations of $\Lambda(Z)$ and $\Lambda(X)$ are given by 
\bea
        \Lambda(Z) = \left( \begin{array}{cccc} 
        1 & 0 & 0 & 0 
\\      0 & 1 & 0 & 0 
\\      0 & 0 & 1 & 0 
\\      0 & 0 & 0 & -1
        \end{array} \right) 
% \\
,\quad
        \Lambda(X) = \left( \begin{array}{cccc} 
        1 & 0 & 0 & 0 
\\      0 & 1 & 0 & 0 
\\      0 & 0 & 0 & 1 
\\      0 & 0 & 1 & 0 
        \end{array} \right) 
.
\eea
We will repeatedly use the following identities involving 
$\Lambda(X)$ and $\Lambda(Z)$: 
\bea 
(I \ms \ot \ms H) \, \Lambda(Z) \, (I \ms \ot \ms H) & = & \Lambda(X) \,,
\label{eq:cxcz}
\\
(H \ms \ot \ms H) \, \Lambda(X) \, (H \ms \ot \ms H) & = & V(X) \,,
\label{eq:cxxc}
\\
\Lambda(Z) \, (X \ms \ot \ms I) \, \Lambda(Z) & = & X \ms \ot \ms Z \,,
\label{eq:czcom1}
\\
\Lambda(Z) \, (Z \ms \ot \ms I) \, \Lambda(Z) & = &  Z \ms \ot \ms I \,, 
\label{eq:czcom2}
\\
\Lambda(X) \, (X \ms \ot \ms I) \, \Lambda(X) & = & X \ms \ot \ms X \,,
\label{eq:cxcom1}
\\
\Lambda(X) \, (I \ms \ot \ms X) \, \Lambda(X) & = & I \ms \ot \ms X \,,
\label{eq:cxcom2}
\\
\Lambda(X) \, (Z \ms \ot \ms I) \, \Lambda(X) & = &  Z \ms \ot \ms I \,, 
\label{eq:cxcom3}
\\
\Lambda(X) \, (I \ms \ot \ms Z) \, \Lambda(X) & = &  Z \ms \ot \ms Z \,.
\label{eq:cxcom4}
\eea
Equation \eq{cxcz} shows that $\Lambda(Z)$ and $\Lambda(X)$ differ only by
the action of single-qubit unitary gates.  Given the ability to
perform single-qubit unitaries, either $\Lambda(Z)$ or $\Lambda(X)$ is
universal for quantum computation.

We only consider {\em projective} measurements, since using
generalized measurements trivializes the problem.  A projective
measurement can be specified by orthogonal subspaces of the measured
Hilbert space; the measurement projects the state onto one subspace
and outputs the subspace label.  Common ways to specify a measurement
include a partition of a basis or the eigenspaces of a Hermitian
operator.  Throughout the paper, a measurement of a Hermitian operator
$O$ is denoted by $M_O$.  

A single-qubit measurement along the computational basis
$\{|0\>,\allowbreak |1\>\}$ is equivalent to $M_Z$.
It has the circuit symbol 
\bea
\setl{0.6mm}
\bp(40,10)
\put(5,5){\line(1,0){5}}
\put(10,0){\fb(14,10){$/$}}
\qbezier(11,2)(17,8)(23,2)
\put(24,7){\line(1,0){7.5}}
\put(24,8){\line(1,0){7.5}}
\put(24,2){\line(1,0){7.5}}
\put(31.5,2){\mb(6,10){$j$}}
\ep
\label{eq:measure}
\eea
Throughout the paper, a double line coming out of a measurement box
represents the classical measurement outcome, and a single line
represents the post-measurement quantum state.
A single-qubit measurement along the basis $\{U^\dg |0\>,\allowbreak
U^\dg |1\>\}$ ($U$ unitary) is equivalent to $M_{U^\dagger Z U}$,
which is equivalent to the sequence of operations $U$, $M_Z$, and
finally $U^\dagger$ on the post-measurement quantum state.
Whenever the post-measurement quantum state is irrelevant, the
measurement is simply depicted as
\bea
\setl{0.6mm}
\bp(42,10)
\put(0,5){\line(1,0){5}}
\put(5,0){\fb(10,10){$U$}}
\put(15,5){\line(1,0){5}}
\put(20,0){\fb(14,10){$/$}}
\qbezier(20,2)(27,8)(34,2)
\put(34,4){\line(1,0){7.5}}
\put(34,6){\line(1,0){7.5}}
\ep
\label{eq:umeasure}
\eea
and conversely, we identify subcircuits of the form given by \eq{umeasure}
as single-qubit measurements.

In addition to complete two-qubit measurements, we will also use {\em
incomplete} two-qubit measurements.  For example, $M_{Z \ot Z}$
measures the parity defined in the computational basis.  As another
example, $M_{U^\dagger Z U \ot Z}$ is equivalent to the sequence of
operations $U \ot I$, $M_{Z \ot Z}$, and $U^\dagger \ot I$.

We will often encounter a measurement of the form
\bea
\setlength{\unitlength}{0.5mm}
\bp(80,22)
\put(5,15){\line(1,0){5}}
\put(5,5){\line(1,0){5}}
\put(10,11){\fb(10,8){$U$}}
\put(10,1){\fb(10,8){$V$}}
\put(20,15){\line(1,0){20}}
\put(20,5){\line(1,0){20}}
\put(30,5){\line(0,1){10}}
\put(28,3){\mb(4,4){$\times$}}
\put(28,13){\mb(4,4){$\times$}}

\put(40,1){\fb(10,8){$V^\dagger$}}
\put(50,5){\line(1,0){5}}
\put(40,11){\fb(10,8){$H$}}
\put(50,15){\line(1,0){5}}

\put(55,11){\framebox(10,8){$/$}}
\qbezier(55,11)(60,18)(65,11)
\put(65,15.5){\line(1,0){5}}
\put(65,14.5){\line(1,0){5}}
\put(70,10){\mb(6,10){$j$}}

\ep
\label{eq:intrigue} 
\eea
where $U,V$ are arbitrary single-qubit gates.
The classical outcome $j$ corresponds to the measurement
of $(U^\dg X U) \ot (V^\dg Z V)$, because 
$(H \ms \ot \ms I) \, \Lambda(Z) \, (U \ms \ot \ms V)$ maps the $\pm 1$
eigenspace of $(U^\dg X U) \ot (V^\dg Z V)$ onto the $\pm 1$
eigenspace of $Z \ot I$.
However, $M_{(U^\dg X U) \ot (V^\dg Z V)}$ does not give rise to the
correct post-measurement quantum state in \eq{intrigue}.
This requires an extra measurement $M_{U^\dg Z U}$ on the first
qubit and an extra gate $(V^\dg Z V)^k$ on the second qubit if the
outcome of $M_{U^\dg Z U}$ is $k$.
In other words, the following circuit is {\em equivalent} to \eq{intrigue}: 
\bea
\setlength{\unitlength}{0.5mm}
\bp(122,32)
\put(3,19){\line(1,0){5}}
\put(3,10){\line(1,0){5}}

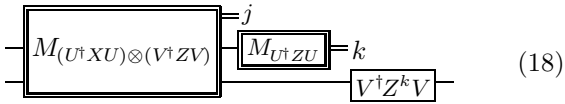
\put(60,27){\line(1,0){5}}
\put(60,28){\line(1,0){5}}
\put(65,25){\mb(5,6){\small $j$}}

\put(60,19){\line(1,0){5}}
\put(60,10){\line(1,0){35}}

\put(8,6){\fb(52,24){}}

\put(9,7){\fb(50,22){$M_{\ms (U^{\! \dg} \ms X \ms U \ms) 
                        \ot (V^{\! \dg} \ms Z \ms V \!)}$}}

\put(65,13.5){\fb(24,10){\small $M_{\ms U^{\! \dg} \ms Z \ms U}$}}
\put(66,14.5){\fb(22,8){}}

\put(89,18.5){\line(1,0){5}}
\put(89,19.5){\line(1,0){5}}
\put(94,14.5){\mb(6,8){$k$}}

\put(95,5){\fb(22,8){\small $V^\dg \! Z^k V \ms$}}
\put(117,10){\line(1,0){5}}

\ep
\label{eq:intrigue2} 
\eea
where double-lined boxes are used to represent the measurements. 

We emphasize that it is useful to view a circuit as an abstract
representation of the evolution of quantum or classical information.
A quantum circuit is often used to represent physical registers and
transformations, but such association is not generally necessary, as
is manifest in our discussion of circuit simulations in the next
section.  

%%%%%%%%%%%%%%%%%%%%%%%%%%%%%%%%%%%%%%%%%%%%%%%%%%%%%%%%%%%%%%%%%%%%%%%%%
\subsection{Circuit simulation with Pauli errors}
\label{sec:circuitsim}

We now describe a notion of circuit simulation useful in the 1WQC and
the TQC models.  Most measurements in the TQC and 1WQC models output
random outcomes and induce Pauli errors that are known functions of
the measurement outcome.
However, the presence of such known errors is not a hindrance to the
computation, provided subsequent measurements are adapted accordingly.
Since our schemes share the same feature, it is useful to introduce
some conventions that simplify later discussions of
simulation.

A quantum circuit consists of ancilla preparations, gates, and
measurements endowed with a partial time-ordering.  We can group
together operations that can be performed in parallel in a time step,
although the grouping may not be unique.
The minimal number of time steps is called the {\em logical depth} of
the circuit.
For each grouping of $\C$, an input state $|\psi_0\>$ specifies a
sequence of quantum states $\{|\psi_i\>\}$ where $|\psi_i\>$ is the
quantum state at the end of the $i$th time step of the computation.
We say that a circuit $\C'$ {\em simulates $\C$ with Pauli errors} if
there is a grouping of $\C$ so that, for any input state $|\psi_0\>$ and
any given Pauli error $P$, applying $\C'$ on the input $P |\psi_0\>$
produces a sequence of states $P_i |\psi_i\>$ where $P_i$ are known
Pauli errors.
These Pauli errors redefine the intermediate states and the final
measurement outcomes, but do not affect the intended computation.
{From} now on, we will simply say that $\C'$ \emph{simulates} $\C$ to
mean that $\C'$ simulates $\C$ with Pauli errors.

Circuit simulation is preserved under the composition of circuits.
Therefore, to simulate a circuit, it suffices to simulate individual
circuit elements.  Although circuit elements may act on part of an
entangled state, it is not difficult to see that it is sufficient to
verify the simulation of a circuit element for all possible pure state
inputs.
Furthermore, universality can be proved by showing how to efficiently
simulate all possible circuit elements required for universal quantum
computation.

In the context of measurement-based quantum computation, initial (or
ancilla) state preparation and read out of computation results can be
simulated as follows.
We will only need initial states that can be prepared by a simple
measurement, up to known Pauli errors.  For example, measuring a
single qubit in the computational basis with outcome $d\in\{0,1\}$
results in the state $X^d |0\>$.  Similarly, a measurement of two
qubits in the Bell basis produces a singlet state up to a known Pauli
error.  Throughout the paper, we suppress the known Pauli errors in
the ancillas whenever their effects are straightforward, so as to keep
the discussion and the simulation circuits simple.  
We also omit physically irrelevant global phase factors that arise
from the composition of Pauli errors.
We restrict our attention to measurements that are deterministically
affected by known Pauli errors, so that the actual outcomes and the
knowledge of the Pauli errors can be used to determine the
Pauli-error-free measurement outcomes.  For example, the result of a
measurement in the computational basis is simply flipped by 
an $X$ error and unaffected by a $Z$ error.  
Now, it suffices to focus on simulating a universal set of unitary
gates in the measurement-based model of quantum computation.

%%%%%%%%%%%%%%%%%%%%%%%%%%%%%%%%%%%%%%%%%%%%%%%%%%%%%%%%%%%%%%%%%%%%%%%%%
\subsection{The TQC model}
\label{sec:tqc}

In this section, we review some elements of the TQC.  Our review
follows the simplified approach of \cite{Leung03t}, which allows the
computation to proceed with a deterministic number of steps.
The universality of the TQC model follows from the ability to
simulate any single-qubit gate $U$ and a two-qubit gate such as
$\Lambda(X)$.

The crucial ingredient of the simulation is
teleportation~\cite{Bennett93}, which transmits a qubit $|\psi\>$
using the following circuit:
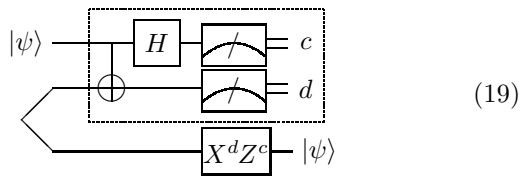
\bea
\setlength{\unitlength}{0.6mm}
\centering
\bp(80,39)
\put(0,25){\makebox(12,10){$|\psi\rangle$}}

\put(12,30){\line(1,0){18}}
\put(12,20){\line(1,0){33}}
\put(5,13){\line(1,1){7}}
\put(5,13){\line(1,-1){7}}

\put(25,30){\line(0,-1){13}}
\put(25,20){\circle{6}}

\put(30,25){\fb(10,10){$H$}}
\put(40,30){\line(1,0){5}}

\put(12, 6){\line(1,0){33}}

\put(45, 1){\framebox(15,10){$X^d \ms Z^c$}}

\put(60, 6){\line(1,0){5}}
\put(65, 1){\makebox(12,10){$|\psi\rangle$}}

\put(45,15){\framebox(14,9){$/$}}
\qbezier(45,17)(52,23)(59,17)
\put(59,21){\line(1,0){5}}
\put(59,19){\line(1,0){5}}
\put(63,15){\mb(10,10){$d$}}

\put(45,25){\framebox(14,9){$/$}}
\qbezier(45,27)(52,33)(59,27)
\put(59,31){\line(1,0){5}}
\put(59,29){\line(1,0){5}}
\put(63,25){\mb(10,10){$c$}}

\put(12,30){\line(1,0){18}}

\put(20,12.5){\dashbox{0.5}(52,25){}}

\ep
\label{eq:teleport} 
\eea
When two qubits are connected on the left side of a circuit diagram,
as the bottom two qubits are in \eq{teleport}, they denote a two-qubit
maximally entangled state $|\Phi_{00}\> = (|00\> + |11\>)/\sqrt 2$.
The dashed box performs a {\em Bell measurement} along the basis
\bea
        |\Phi_{00}\> &=& \os2 (|00\> + |11\>) \,,~~
        |\Phi_{10}\>  =  \os2 (|00\> - |11\>) \,,
\non\\
        |\Phi_{01}\> &=& \os2 (|01\> + |10\>) \,,~~
        |\Phi_{11}\>  =  \os2 (|01\> - |10\>) \,.
\non
\eea
The teleportation circuit can be verified by rewriting the initial
state $|\psi\> |\Phi_{00}\>$ as $\frac{1}{2} \sum_{c,d} |\Phi_{cd}\> 
\ot (Z^{c} X^{d} |\psi\>)$.

The teleportation circuit \eq{teleport} simulates the identity gate
(in the sense described in \sec{circuitsim}).  In fact, it does so
even when the post-measurement correction $X^d Z^c$ is omitted: for
any input state $X^a Z^b |\psi\>$, the output from the teleportation
circuit without correction is simply $X^{a+d} Z^{b+c} |\psi\>$ (up to
a global phase).

Suppose we want to simulate a single-qubit gate $U$.
Consider a circuit in which we apply the gate $U' = U Z^b X^a$ to the
input state $X^a Z^b \hspace*{0.1ex} |\psi\>$ and perform teleportation on the
resulting state $U|\psi\>$ without the correction.  Following
\eq{teleport}, the output quantum state of the circuit is $X^d Z^c U
|\psi\>$.
% , up to an unimportant global phase factor.  
In other words, the following circuit simulates $U$:
\bea
\setlength{\unitlength}{0.6mm}
\centering
\bp(80,37)
\put(-22.5,23){\makebox(20,10){$X^a Z^b \hspace*{0.1ex} |\psi\rangle$}}

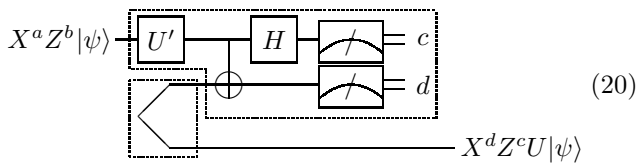
\put(0,28){\line(1,0){5}}
\put(5,23){\fb(10,10){$U'$}}

\put(15,28){\line(1,0){15}}
\put(12,18){\line(1,0){33}}
\put(5,11){\line(1,1){7}}
\put(5,11){\line(1,-1){7}}

\put(25,28){\line(0,-1){13}}
\put(25,18){\circle{6}}

\put(30,23){\fb(10,10){$H$}}
\put(40,28){\line(1,0){5}}

\put(12,4){\line(1,0){63}}

\put(75,-1){\mb(30,10){$X^d \ms Z^c U |\psi\rangle$}}

\put(45,13){\framebox(14,9){$/$}}
\qbezier(45,15)(52,21)(59,15)
\put(59,19){\line(1,0){5}}
\put(59,17){\line(1,0){5}}
\put(63,13){\mb(10,10){$d$}}

\put(45,23){\framebox(14,9){$/$}}
\qbezier(45,25)(52,31)(59,25)
\put(59,29){\line(1,0){5}}
\put(59,27){\line(1,0){5}}
\put(63,23){\mb(10,10){$c$}}

\multiput(20,10.5)(1,0){51}{\line(1,0){0.5}}
\multiput(20,10.5)(0,1){11}{\line(0,1){0.5}}
\multiput(20,21.5)(-1,0){17}{\line(-1,0){0.5}}
\multiput(3,21.5)(0,1){13}{\line(0,1){0.5}}
\multiput(3,34.5)(1,0){68}{\line(1,0){0.5}}
\multiput(71,10.5)(0,1){24}{\line(0,1){0.5}}

\put(3,2){\dashbox{0.5}(15,17){}}

\ep
\label{eq:teleportu} 
\eea
The circuit in~\eq{teleportu} can be divided into subcircuits, each of
which can be simulated by a single two-qubit measurement.
The first subcircuit (the dashed box in the lower left corner) is just
the preparation of the state $|\Phi_{00}\>$. 
It can be simulated by a single Bell measurement, up to a Pauli error.
The rest of the circuit, $(U' \ot I)$ followed by a Bell measurement,
is just a two-qubit measurement along a rotated Bell basis $\{ (U'^\dg
\ot I) |\Phi_{cd}\> \}$. 
Composing these two measurements provides a simulation of
\eq{teleportu}, and thus a simulation of $U$, in the TQC model.

An alternative simulation is applicable to gates in the
Clifford group \cite{Gottesman99t}:
\bea
\setlength{\unitlength}{0.6mm}
\centering
\bp(80,37)
\put(-22.5,23){\makebox(20,10){$X^a Z^b \hspace*{0.1ex} |\psi\rangle$}}

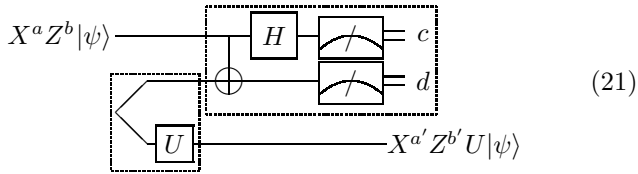
\put(0,28){\line(1,0){15}}

\put(15,28){\line(1,0){15}}
\put(7,18){\line(1,0){38}}
\put(0,11){\line(1,1){7}}
\put(0,11){\line(1,-1){7}}

\put(25,28){\line(0,-1){13}}
\put(25,18){\circle{6}}

\put(30,23){\fb(10,10){$H$}}
\put(40,28){\line(1,0){5}}

\put(7,4){\line(1,0){2}}
\put(9,0){\fb(8,8){$U$}}
\put(17,4){\line(1,0){43}}

\put(60,-1){\mb(30,10){$X^{\ms a'} \ms Z^{\ms b'} U |\psi\rangle$}}

\put(45,13){\framebox(14,9){$/$}}
\qbezier(45,15)(52,21)(59,15)
\put(59,19){\line(1,0){5}}
\put(59,17){\line(1,0){5}}
\put(63,13){\mb(10,10){$d$}}

\put(45,23){\framebox(14,9){$/$}}
\qbezier(45,25)(52,31)(59,25)
\put(59,29){\line(1,0){5}}
\put(59,27){\line(1,0){5}}
\put(63,23){\mb(10,10){$c$}}

\put(20,10.5){\dashbox{0.5}(51,24){}}
\put(-1,-2){\dashbox{0.5}(19.5,21.5){}}

\ep
\label{eq:teleportgc} 
\eea
The effect of the above circuit is to teleport the input state $X^a
Z^b |\psi\>$ (without the correction) to produce the state $X^{a+d}
Z^{b+c} |\psi\>$,
% (up to an unimportant global phase factor), 
and then to apply the gate $U$.  The output state is thus $U X^{a+d}
Z^{b+c} |\psi\>$.  Using the fact $U$ is in the Clifford group, the
output state can be rewritten as $X^{\ms a'} Z^{\ms b'} U |\psi\>$ for
known $a',b'$.

The simulation of two-qubit gates is similar to that depicted in
\eq{teleportu} and \eq{teleportgc}.
We omit the details of the existing simulation schemes, and we refer
the interested reader to \cite{Leung03t}.  Instead, we will present a
simplified simulation in \sec{newtqc}.

Comparing \eq{teleportu} and \eq{teleportgc}, the latter simulation
has a simpler teleportation measurement but a more complicated initial
state $(I \ot U) |\Phi_{00}\>$.
This tradeoff is useful in the
simulation of two-qubit gates in the Clifford group.

Note that in the TQC model we assume the ability to
apply measurements to any subset of the qubits, without worrying about
the respective locations of those qubits, just as in the circuit
model.  The TQC is simply a method for implementing each gate in the
circuit model by a sequence of measurements.

%%%%%%%%%%%%%%%%%%%%%%%%%%%%%%%%%%%%%%%%%%%%%%%%%%%%%%%%%%%%%%%%%%%%%%%%%
\subsection{The one-way quantum computer model}
\label{sec:1wqcmodel}

Since we will present a simple systematic derivation of 1WQC-like
schemes, we omit the details of the existing 1WQC schemes, and refer
interested readers to the ingenious schemes discussed in
\cite{Raussendorf01a,Raussendorf02a,Raussendorf03}.
Instead, we focus on features of the 1WQC that differ from the TQC
model.

The 1WQC is based on an input state known as the {\em cluster state}
\cite{Raussendorf00}.  The cluster state consists of a two-dimensional
square lattice of qubits.  To simulate a computation of $n$ qubits
with logical depth $m$, the lattice is chosen to be of size $O(n)
\times O(m)$.
One way of defining the cluster state is as the result of a two-stage
preparation procedure: ($i$) prepare all lattice sites in the state
$|+\>=\sos2(|0\>+|1\>)$, and ($ii$) apply $\Lambda(Z)$ between each pair of
qubits that are adjacent in the lattice. Since the $\Lambda(Z)$ operations
all commute, it does not matter in what order they are applied.  Note
that this prescription is merely a convenient way of defining the
cluster state, and there is no need to actually prepare it by
following these steps. 

The cluster states naturally generalize to {\em graph states}, for
which an arbitrary graph defines the adjacency relation
\cite{Werner01,Raussendorf03,Briegel03}.
We will use a feature of any graph state called the {\em deletion
principle}.  
When one of the qubits is measured in the computational basis, the
unmeasured qubits will be left in a different graph state (up to known
Pauli errors) corresponding to the graph obtained by deleting the
measured qubit from the original graph.
In the 1WQC model, once the cluster state is prepared, an {\em
arbitrary} circuit $\C$ can be simulated using only single-qubit
measurements.  The first step is to ``imprint'' the circuit $\C$ onto
the cluster state by deleting qubits from the lattice to obtain some
graph state that depends on $\C$.  Roughly speaking, this graph
consists of horizontal lines of vertices, each line corresponding to a
qubit in $\C$, and vertical edges connecting neighboring lines
corresponding to interactions between the simulated qubits.
The qubits in the graph are then measured from the left to the right.
Each measurement propagates quantum information from the measured
qubit to its unmeasured right neighbor.
In general, the measurement basis will depend on previous measurement
outcomes.
Various regions of the graph simulate subcircuits.  Qubits entering a
region from the left boundary carry the input state of the circuit
element, and qubits exiting at the right boundary carry the output. 
When composing element-wise simulation in the 1WQC model, the output
registers of one region have to match the input registers of the next
region.
A schematic diagram for a computation in the 1WQC model is given 
in Fig.~1. 
% \ref{fig:rb}
% 
%\begin{figure}[ht]
%\includegraphics[width=3in]{rbnew}
% \caption{A schematic diagram of a 1WQC computation.
% Figure courtesy H.\ J.\ Briegel and R.\ Raussendorf.}
%\label{fig:rb}
%\end{figure}

Simulation of a circuit using the 1WQC is discussed in detail in
\cite{Raussendorf02a}.  The precise protocols and their verification
are beyond the scope of this paper.  Interpretations in terms of
valence bond solids \cite{Verstraete03} and teleportation
\cite{Aliferis04,Nielsen03rb} have been reported recently.
In this paper, the derivation is based on a different underlying
principle and the schemes are significantly simpler than those
discussed in~\cite{Raussendorf02a}.

%%%%%%%%%%%%%%%%%%%%%%%%%%%%%%%%%%%%%%%%%%%%%%%%%%%%%%%%%%%%%%%%%%%%%%%%%
\subsection{The substrate representation} 

Circuit representations of a computation in the 1WQC model turn out to
be rather unwieldy.  Instead, we prefer to use a more concise
``substrate representation'' in which we label each vertex in the
graph representing the cluster with the measurement to be performed.
The measurement bases may depend on previous measurement outcomes,
and this dependency should be indicated in the labeling.  Note that
the interdependence of the measurement outcomes and measurement bases
specifies a partial order in which the qubits must be measured.  An example
of the substrate representation is~(\ref{eq:1bitsub}), in
\sec{1bit-telep}, which simulates the circuit
in~(\ref{eq:circuiteg}).

The substrate representation for simulating a certain circuit often
visually resembles the simulated circuit (cf.~(\ref{eq:circuiteg})).
One can identify the different physical qubits that carry the
propagating quantum state at different times with propagation of
quantum information in time, and the various regions of the graph
state corresponding to different quantum gates with the action of
those gates themselves.

%%%%%%%%%%%%%%%%%%%%%%%%%%%%%%%%%%%%%%%%%%%%%%%%%%%%%%%%%%%%%%%%%%%%%%
\section{One-bit teleportation and simple simulation circuits}

\label{sec:1btsims}

All the measurement-based models of quantum computation share the
common feature that measurements are not performed solely on the
qubits storing the data.  The reason is that doing so would destroy
the coherence essential to quantum computation.  Instead, ancilla
qubits are prepared, and then measurements are used to interact the
data with the ancilla.  By choosing the measurements and initial
states of the ancilla carefully, we can ensure that coherence is
preserved.  Even more remarkably, with suitable choices of ancilla and
measurements, it is possible to effect a universal set of quantum
gates.

In this section, we introduce two circuits that offer perhaps the
simplest example of these principles in action, the one-bit
teleportation circuits introduced in~\cite{Zhou00}.  We will show how
the one-bit teleportation circuits can be used to derive simulation
circuits for a universal set of quantum gates.  These simulation
circuits are, again, based solely on measurements and interaction with
an ancilla.  In later sections, we will see that these simulation
circuits can be used to derive all known measurement-based models
of quantum computation.

The one-bit teleportation circuits are as follows~\cite{Zhou00}:
\bea
\setlength{\unitlength}{0.6mm}
\bp(80,33)
\put(-15,22.5){\mb(40,10){$Z$-teleportation:}}
\put(-2,10){\makebox(12,10){$|\psi\rangle$}}
\put(0,0){\makebox(10,10){$|0\>$}}

\put(10,15){\line(1,0){20}}
\put(10,5){\line(1,0){30}}

\put(25,15){\line(0,-1){13}}
\put(25,5){\circle{6}}

\put(30,11){\fb(8,8){$H$}}
\put(38,15){\line(1,0){7}}

\put(35, 0){\makebox(30,10){$Z^c |\psi\rangle$}}

\put(45,10){\framebox(14,9){$/$}}
\qbezier(45,12)(52,18)(59,12)
\put(59,16){\line(1,0){5}}
\put(59,14){\line(1,0){5}}
\put(63,10){\mb(10,10){$c$}}

\ep
\label{eq:zt} 
\\
\setlength{\unitlength}{0.6mm}
\bp(80,33)
\put(-15,22.5){\mb(40,10){$X$-teleportation:}}
\put(-2,10){\makebox(12,10){$|\psi\rangle$}}
\put(0,0){\makebox(10,10){$|0\>$}}

\put(10,15){\line(1,0){35}}
\put(10,5){\line(1,0){5}}

\put(30,18){\line(0,-1){13}}
\put(30,15){\circle{6}}

\put(15,1){\fb(8,8){$H$}}
\put(23,5){\line(1,0){17}}

\put(35, 0){\makebox(30,10){$X^d |\psi\rangle$}}

\put(45,10){\framebox(14,9){$/$}}
\qbezier(45,12)(52,18)(59,12)
\put(59,16){\line(1,0){5}}
\put(59,14){\line(1,0){5}}
\put(63,10){\mb(10,10){$d$}}
\ep
\label{eq:xt} 
\eea
These circuits are analogous to teleportation in that they move a qubit from
one register to another. 
The circuits are named after the Pauli corrections required to fully
recover the input state.
The circuits are easily verified \cite{Zhou00}.
Note that the $Z$- and $X$-teleportation circuits can be
interconverted by using the input state $H|\psi\>$ and applying
\eqs{hadamard}{cxxc}.
Moreover, by rewriting the state $|\Phi_{00}\>$ as $\Lambda(X) (H \ot
I) |00\>$, the teleportation circuit in \eq{teleport} can be viewed as
a composition of a $Z$-teleportation followed by an $X$-teleportation.
Thus, all results derived from \eq{teleport} and \eq{zt}--\eq{xt} can
be derived from one of \eq{zt} or \eq{xt} alone.

We now derive from \eq{zt} and \eq{xt} some useful simulation circuits
for a universal set of gates.  In subsequent sections, we will
systematically derive schemes in the TQC and the 1WQC models using
these simulation circuits.

Consider procedures analogous to the simulation circuits
\eq{teleportu}--\eq{teleportgc}, but based on one-bit teleportation.
To simulate a single-qubit gate $U$ acting on an input state $|\psi\>$
with Pauli error $X^a Z^b$, a simulation circuit can consist of first
applying $U'= U Z^b X^a$ before either form of one-bit teleportation:
\bea
\setlength{\unitlength}{0.6mm}
\bp(80,20)
\put(-13,10){\mb(20,10){$X^a Z^b |\psi\rangle$}}
\put(0,0){\makebox(10,10){$|0\>$}}

\put(10,15){\line(1,0){4}}
\put(14,11){\fb(8,8){$U'$}}
\put(22,15){\line(1,0){8}}
\put(10,5){\line(1,0){55}}

\put(26,15){\line(0,-1){13}}
\put(26,5){\circle{6}}

\put(30,11){\fb(8,8){$H$}}
\put(38,15){\line(1,0){3}}

\put(60, 0){\makebox(30,10){$Z^c U |\psi\rangle$}}

\put(41,11){\framebox(12,8){$/$}}
\qbezier(41,12)(47,18)(53,12)
\put(53,15.5){\line(1,0){4}}
\put(53,14.5){\line(1,0){4}}
\put(55,10){\mb(10,10){$c$}}

\put(12,0){\dashbox{0.5}(50,21){}}

\ep
\label{eq:uzt} 
\eea
\bea
\setlength{\unitlength}{0.6mm}
\bp(80,20)
\put(-13,10){\mb(20,10){$X^a Z^b |\psi\rangle$}}
\put(0,0){\makebox(10,10){$|0\>$}}

\put(10,15){\line(1,0){5}}
\put(10,5){\line(1,0){5}}

\put(30,18){\line(0,-1){13}}
\put(30,15){\circle{6}}

\put(15,1){\fb(8,8){$H$}}
\put(23,5){\line(1,0){42}}

\put(15,11){\fb(8,8){$U'$}}
\put(23,15){\line(1,0){18}}

\put(60, 0){\makebox(30,10){$X^d U |\psi\rangle$}}

\put(41,11){\framebox(12,8){$/$}}
\qbezier(41,12)(47,18)(53,12)
\put(53,15.5){\line(1,0){4}}
\put(53,14.5){\line(1,0){4}}
\put(55,10){\mb(10,10){$d\,$}}

\multiput(12,20.5)(1,0){50}{\line(1,0){0.5}}
\multiput(62,20.5)(0,-1){21}{\line(0,-1){0.5}}
\multiput(62,-0.5)(-1,0){37}{\line(-1,0){0.5}}
\multiput(25,-0.5)(0,1){11}{\line(0,1){0.5}}
\multiput(25,10)(-1,0){13}{\line(-1,0){0.5}}
\multiput(12,10)(0,1){11}{\line(0,1){0.5}}

\ep
\label{eq:uxt} 
\eea

We will see that \eq{uzt} and \eq{uxt} are most useful in the TQC
model.  In the 1WQC model, more specialized simulation circuits are
required.
To simulate the rotation $Z_\t$, our simulation circuit takes the
input state $X^a Z^b |\psi\>$ and applies $Z_{(-1)^a \t}$, followed by
$Z$-teleportation.  When the measurement outcome is $c$, the output
state is $Z^c Z_{(-1)^a \t} X^a Z^b |\psi\>$.  Using the identity $X^a
Z_{(-1)^a \t} X^a = Z_{\t}$, the output state is $X^a Z^{b+c} Z_\t
|\psi\>$. 
%, up to an unimportant global phase.  
This is summarized in the circuit
\bea
\setlength{\unitlength}{0.6mm}
\bp(80,20)
\put(-14,10){\mb(22,10){$X^a Z^b|\psi\rangle$}}
\put(0,0){\makebox(10,10){$|0\>$}}

\put(10,15){\line(1,0){10}}
\put(10,5){\line(1,0){30}}

\put(15,15){\line(0,-1){13}}
\put(15,5){\circle{6}}

\put(20,11){\fb(16,8){\footnotesize $Z_{(\!-\!1)^{\!a} \t}$}}
\put(36,15){\line(1,0){2}}

\put(38,11){\fb(6,8){$H$}}
\put(44,15){\line(1,0){2}}

\put(19,10){\dashbox{0.5}(51,10){}}

\put(35, 0){\makebox(50,10){$X^a Z^{b+c} Z_\t |\psi\rangle$}}

\put(46,11){\framebox(12,8){$/$}}
\qbezier(46,12)(52,18)(58,12)
\put(58,16){\line(1,0){5}}
\put(58,14){\line(1,0){5}}
\put(62,10){\mb(10,10){$c$}}

\ep
\label{eq:zrot} 
\eea
where we have commuted $\Lambda(X)$ and $Z_{(-1)^a \t}$. 
Similarly, for the gate $X_\t$, consider a simulation circuit with
an input state $X^a Z^b |\psi\>$, a gate $X_{(-1)^b \t}$ applied to
the input, followed by $X$-teleportation.  When the measurement
outcome is $d$, the output state is $X^d X_{(-1)^b \t} X^a Z^b |\psi\>
= X^{a+d} Z^{b} X_\t |\psi\>$.  This is summarized in the circuit
\bea
\setlength{\unitlength}{0.6mm}
\centering
\bp(80,20)
\put(-14,10){\mb(22,10){$X^a Z^b|\psi\rangle$}}
\put(0,0){\makebox(10,10){$|0\>$}}

\put(10,15){\line(1,0){18}}
\put(10,5){\line(1,0){3}}

\put(22,18){\line(0,-1){13}}
\put(22,15){\circle{6}}

\put(13,1){\fb(6,8){$H$}}
\put(19,5){\line(1,0){17}}

\put(28,11){\fb(16,8){\footnotesize $X_{(\!-\!1)^{\!b} \t}$}}
\put(44,15){\line(1,0){2}}

\put(27,10){\dashbox{0.5}(43,10){}}

\put(35, 0){\makebox(50,10){$X^{a+d} Z^{b} X_\t |\psi\rangle$}}

\put(46,11){\framebox(12,8){$/$}}
\qbezier(46,12)(52,18)(58,12)
\put(58,16){\line(1,0){5}}
\put(58,14){\line(1,0){5}}
\put(62,10){\mb(10,10){$d$}}

\ep
\label{eq:xrot} 
\eea
where we have commuted $\Lambda(X)$ and $X_{(-1)^b \t}$. 

Finally, we consider a simulation circuit for $\Lambda(Z)$ in which two
$X$-teleportation circuits (without correction) are applied to the
two-qubit input $X^{a_1} Z^{b_1} \ot X^{a_2} Z^{b_2} |\psi\>$,
followed by applying $\Lambda(Z)$: 
\bea
\setlength{\unitlength}{0.6mm}
\bp(80,50)
\put(-12,46){\mb(42,8){\scriptsize 
$X^{\ms a_{\ms 1}} \ms Z^{\ms b_{\ms 1}} \! \ot \! 
 X^{\ms a_{\ms 2}} \ms Z^{\ms b_{\ms 2}} |\psi\rangle$}}
\put(-8,0){\makebox(10,10){$|0\>$}}
\put(-5,47){\vector(1,-3){4.5}}

\put(2,30){\line(1,0){30}}
\put(2,5){\line(1,0){3}}
\put(11,5){\line(1,0){30}}

\put(20,33){\line(0,-1){28}}
\put(20,30){\circle{6}}

\put(5,11){\fb(6,8){$H$}}

\put(32,26){\framebox(10,8){$/$}}
\qbezier(32,26)(37,33)(42,26)
\put(42,30.5){\line(1,0){5}}
\put(42,29.5){\line(1,0){5}}
\put(47,26){\mb(10,8){$d_2$}}

%---------------------------------------------------------------
\put(-8,10){\makebox(10,10){$|0\>$}}

\put(2,40){\line(1,0){30}}
\put(2,15){\line(1,0){3}}
\put(11,15){\line(1,0){30}}

\put(15,43){\line(0,-1){28}}
\put(15,40){\circle{6}}

\put(5,1){\fb(6,8){$H$}}

\put(32,36){\framebox(10,8){$/$}}
\qbezier(32,36)(37,43)(42,36)
\put(42,40.5){\line(1,0){5}}
\put(42,39.5){\line(1,0){5}}
\put(47,36){\mb(10,8){$d_1$}}

%-------------------------------------------------------------
\put(27,15){\line(0,-1){10}}
\put(25,13){\mb(4,4){$\times$}}
\put(25,3){\mb(4,4){$\times$}}

\put(40,5){\mb(60,10){\scriptsize
$ (X^{\ms a_{\ms 1}'} \ms Z^{\ms b_{\ms 1}'} \! \ot \! 
 X^{\ms a_{\ms 2}'} \ms Z^{\ms b_{\ms 2}'} ) 
\Lambda(Z) |\psi\rangle$}}

\ep
\label{eq:xtcz} 
\eea
When the measurement outcomes of the two $X$-teleportation steps are
$d_1$ and $d_2$, the output state of the circuit is 
$ \Lambda(Z) (X^{\ms a_{\ms 1} + d_{\ms 1}} \ms Z^{\ms b_{\ms 1}} \! \ot \!
         X^{\ms a_{\ms 2} + d_{\ms 2}} \ms Z^{\ms b_{\ms 2}}) 
|\psi\rangle$.  
Using \eq{czcom1} and \eq{czcom2}, the output state is
$(X^{\ms a_{\ms 1} + d_{\ms 1}} \ms 
Z^{\ms b_{\ms 1} + a_{\ms 2} + d_{\ms 2}} \! \ot \!
X^{\ms a_{\ms 2} + d_{\ms 2}} \ms 
Z^{\ms b_{\ms 2} + a_{\ms 1} + d_{\ms 1}}) \Lambda(Z) |\psi\rangle$.  
Thus in \eq{xtcz}, $a_1' = a_{\ms 1} + d_{\ms 1}$, $b_1' = b_{\ms 1} +
a_{\ms 2} + d_{\ms 2}$, $a_2' = a_{\ms 2} + d_{\ms 2}$, and $b_2' =
b_{\ms 2} + a_{\ms 1} + d_{\ms 1}$.

We can derive useful simulation circuits from \eq{xtcz}.  Suppose we
commute $\Lambda(Z)$ to the left of the controlled-{\sc not}s, and reorder
the qubits so that the second qubit from the top becomes the last:
\bea
\setlength{\unitlength}{0.6mm}
\bp(80,55)
\put(-12,46){\mb(42,8){\scriptsize 
$X^{\ms a_{\ms 1}} \ms Z^{\ms b_{\ms 1}} \! \ot \!  
 X^{\ms a_{\ms 2}} \ms Z^{\ms b_{\ms 2}} |\psi\rangle$}}

\put(-8,47){\line(0,-1){47}}
\put(-8,0){\vector(1,0){8}}
\put(-6,47){\vector(1,-1){7}}

\put(2,0){\line(1,0){30}}
\put(25,-3){\line(0,1){18}}
\put(25,0){\circle{6}}

\put(32,-4){\framebox(10,8){$/$}}
\qbezier(32,-4)(37,3)(42,-4)
\put(42,0.5){\line(1,0){5}}
\put(42,-0.5){\line(1,0){5}}
\put(47,-4){\mb(10,8){$d_2$}}
%---------------------------------------------------------------
\put(11,15){\line(1,0){30}}
\put(2,15){\line(1,0){3}}
\put(-8,10){\makebox(10,10){$|0\>$}}
\put(5,21){\fb(6,8){$H$}}
%---------------------------------------------------------------
\put(5,11){\fb(6,8){$H$}}
\put(-8,20){\makebox(10,10){$|0\>$}}
\put(2,25){\line(1,0){3}}
\put(11,25){\line(1,0){30}}
%---------------------------------------------------------------

\put(2,40){\line(1,0){30}}

\put(25,43){\line(0,-1){18}}
\put(25,40){\circle{6}}

\put(32,36){\framebox(10,8){$/$}}
\qbezier(32,36)(37,43)(42,36)
\put(42,40.5){\line(1,0){5}}
\put(42,39.5){\line(1,0){5}}
\put(47,36){\mb(10,8){$d_1$}}

%-------------------------------------------------------------
\put(17,25){\line(0,-1){10}}
\put(15,23){\mb(4,4){$\times$}}
\put(15,13){\mb(4,4){$\times$}}

\put(40,15){\mb(60,10){\scriptsize
$ (X^{\ms a_{\ms 1}'} \ms Z^{\ms b_{\ms 1}'} \! \ot \! 
 X^{\ms a_{\ms 2}'} \ms Z^{\ms b_{\ms 2}'} ) 
\Lambda(Z) |\psi\rangle$}}

\put(-7,10){\dashbox(28,20){}}

\ep
\label{eq:xtcz2} 
\eea
\hfill\\ Furthermore, for the same input state, the following circuits
produce the same outcomes and corresponding post-measurement states:
\bea
\setlength{\unitlength}{0.6mm}
\bp(105,19)
% \bp(105,20)
% \put(0,0){\fb(105,18){}}
\put(5,5){\line(1,0){32}}
\put(5,15){\line(1,0){17}}
\put(12.5,18){\line(0,-1){13}}
\put(12.5,15){\circle{6}}
\put(22,11){\framebox(10,8){$/$}}
\qbezier(22,11)(27,18)(32,11)
\put(32,15.5){\line(1,0){5}}
\put(32,14.5){\line(1,0){5}}

\put(37,10){\mb(6,10){$j$}}

\put(50,5){\mb(6,10){$=$}}

\put(97,0){\mb(6,10){$j$}}

\put(65,15){\line(1,0){18}}
\put(83,11){\fb(8,8){$X^{\ms j}$}}

\put(91,15){\line(1,0){6}}

\put(65,5){\line(1,0){17}}
\put(72.5,15){\line(0,-1){13}}
\put(72.5,5){\circle{6}}
\put(82,1){\framebox(10,8){$/$}}
\qbezier(82,1)(87,8)(92,1)
\put(92,5.5){\line(1,0){5}}
\put(92,4.5){\line(1,0){5}}
\ep
\label{eq:zzmeas} 
\eea
Thus \eq{xtcz2} implies the following: 
\bea
\setlength{\unitlength}{0.6mm}
% \bp(80,55)
\bp(80,54)
\put(-12,46){\mb(42,8){\scriptsize 
$X^{\ms a_{\ms 1}} \ms Z^{\ms b_{\ms 1}} \! \ot \!  
 X^{\ms a_{\ms 2}} \ms Z^{\ms b_{\ms 2}} |\psi\rangle$}}

\put(-10,47){\line(0,-1){47}}
\put(-10,0){\vector(1,0){8}}
\put(-8,47){\vector(1,-1){7}}

\put(2,0){\line(1,0){45}}
\put(25,0){\line(0,1){18}}
\put(25,15){\circle{6}}

\put(32,11){\framebox(10,8){$/$}}
\qbezier(32,11)(37,18)(42,11)
\put(42,15.5){\line(1,0){5}}
\put(42,14.5){\line(1,0){5}}
\put(47,11){\mb(10,8){$d_2$}}
%---------------------------------------------------------------
\put(11,15){\line(1,0){21}}
\put(2,15){\line(1,0){3}}
\put(-8,10){\makebox(10,10){$|0\>$}}
\put(5,21){\fb(6,8){$H$}}
%---------------------------------------------------------------
\put(5,11){\fb(6,8){$H$}}
\put(-8,20){\makebox(10,10){$|0\>$}}
\put(2,25){\line(1,0){3}}
\put(11,25){\line(1,0){21}}
%---------------------------------------------------------------

\put(2,40){\line(1,0){45}}

\put(25,40){\line(0,-1){18}}
\put(25,25){\circle{6}}

\put(32,21){\framebox(10,8){$/$}}
\qbezier(32,21)(37,28)(42,21)
\put(42,25.5){\line(1,0){5}}
\put(42,24.5){\line(1,0){5}}
\put(47,21){\mb(10,8){$d_1$}}

%-------------------------------------------------------------
\put(17,25){\line(0,-1){10}}
\put(15,23){\mb(4,4){$\times$}}
\put(15,13){\mb(4,4){$\times$}}

\put(40,45){\mb(60,10){\scriptsize
$ (X^{\ms a_{\ms 1}} \ms Z^{\ms b_{\ms 1}'} \! \ot \! 
 X^{\ms a_{\ms 2}} \ms Z^{\ms b_{\ms 2}'} ) 
\Lambda(Z) |\psi\rangle$}}

\put(-8,10){\dashbox(28,20){}}

\put(60,47){\line(0,-1){47}}
\put(60,0){\vector(-1,0){8}}
\put(58,47){\vector(-1,-1){7}}

\ep
\label{eq:xtcz3} 
\eea 
where, according to \eq{zzmeas}, the output $X$ errors in \eq{xtcz3}
are obtained by adding $d_1,d_2$ to $a_1',a_2'$ defined in \eq{xtcz2}.
The results are simply $a_1,a_2$.
Finally, rewrite both controlled-{\sc not}s using \eq{cxcz}, and note
that the state in the dashed box in \eq{xtcz3} is stabilized by $H \ot
H$, giving a ``remote $\Lambda(Z)$'' construction:
\bea
\setlength{\unitlength}{0.6mm}
\bp(80,55)
\put(-12,46){\mb(42,8){\scriptsize 
$X^{\ms a_{\ms 1}} \ms Z^{\ms b_{\ms 1}} \! \ot \!  
 X^{\ms a_{\ms 2}} \ms Z^{\ms b_{\ms 2}} |\psi\rangle$}}

\put(-10,47){\line(0,-1){47}}
\put(-10,0){\vector(1,0){8}}
\put(-8,47){\vector(1,-1){7}}

\put(2,0){\line(1,0){45}}
\put(25,0){\line(0,1){15}}
\put(23,13){\mb(4,4){$\times$}}
\put(23,-2){\mb(4,4){$\times$}}

\put(32,11){\fb(6,8){$H$}}
\put(38,15){\line(1,0){4}}

\put(42,11){\framebox(10,8){$/$}}
\qbezier(42,11)(47,18)(52,11)
\put(52,15.5){\line(1,0){5}}
\put(52,14.5){\line(1,0){5}}
\put(57,11){\mb(10,8){$d_2$}}
%---------------------------------------------------------------
\put(11,15){\line(1,0){21}}
\put(2,15){\line(1,0){3}}
\put(-8,10){\makebox(10,10){$|0\>$}}
\put(5,21){\fb(6,8){$H$}}
%---------------------------------------------------------------
\put(5,11){\fb(6,8){$H$}}
\put(-8,20){\makebox(10,10){$|0\>$}}
\put(2,25){\line(1,0){3}}
\put(11,25){\line(1,0){21}}
%---------------------------------------------------------------

\put(2,40){\line(1,0){45}}
 
\put(25,25){\line(0,1){15}}
\put(23,23){\mb(4,4){$\times$}}
\put(23,38){\mb(4,4){$\times$}}

\put(32,21){\fb(6,8){$H$}}
\put(38,25){\line(1,0){4}}

\put(42,21){\framebox(10,8){$/$}}
\qbezier(42,21)(47,28)(52,21)
\put(52,25.5){\line(1,0){5}}
\put(52,24.5){\line(1,0){5}}
\put(57,21){\mb(10,8){$d_1$}}

%-------------------------------------------------------------
\put(17,25){\line(0,-1){10}}
\put(15,23){\mb(4,4){$\times$}}
\put(15,13){\mb(4,4){$\times$}}

\put(40,45){\mb(60,10){\scriptsize
$ (X^{\ms a_{\ms 1}} \ms Z^{\ms b_{\ms 1}'} \! \ot \! 
 X^{\ms a_{\ms 2}} \ms Z^{\ms b_{\ms 2}'} ) 
\Lambda(Z) |\psi\rangle$}}

\put(65,47){\line(0,-1){47}}
\put(65,0){\vector(-1,0){8}}
\put(63,47){\vector(-1,-1){7}}

\put(-8,10){\dashbox(28,20){}}
\ep
\label{eq:xtcz4} 
\eea
If we perform a remote controlled-{\sc not} by performing $H$ before
and after the remote $\Lambda(Z)$ according to \eq{cxcz}, we obtain the
well-known remote $\Lambda(X)$ circuit due to Gottesman
\cite{Gottesman98h}.
The current derivation is only based on the principle of performing the
desired gate after one-bit teleportation, and is different from the
derivation in \cite{Zhou00}. 

Our last simulation circuit for $\Lambda(Z)$ uses the standard (and
easily-verified) result that the following circuit implements $M_{Z
\ot Z}$ on the two input qubits:
\bea
\setlength{\unitlength}{0.6mm}
\bp(110,35)
\put(0,25){\mb(10,10){$|0\>$}}
\put(10,30){\line(1,0){5}}
\put(15,26){\fb(8,8){$H$}}
\put(23,30){\line(1,0){17}}
\put(40,26){\fb(8,8){$H$}}
\put(48,30){\line(1,0){5}}
\put(53,26){\framebox(10,8){$/$}}
\qbezier(53,26)(58,32)(63,26)
\put(63,30.5){\line(1,0){3}}
\put(63,29.5){\line(1,0){3}}
\put(66,25){\mb(6,10){$j$}}

\put(2,15){\line(1,0){66}}
\put(2,5){\line(1,0){66}}

\put(29,30){\line(0,-1){15}}
\put(27,13){\mb(4,4){$\times$}}
\put(27,28){\mb(4,4){$\times$}}

\put(34,30){\line(0,-1){25}}
\put(32,3){\mb(4,4){$\times$}}
\put(32,28){\mb(4,4){$\times$}}

\put(71,13){\mb(10,10){$=$}}

\put(83,14){\line(1,0){5}}
\put(83,5){\line(1,0){5}}
\put(89,2){\fb(15,20){$M_{\!Z\!\ot\!Z}$}}
\put(88,1){\fb(17,22){}}
\put(105,14){\line(1,0){5}}
\put(105,5){\line(1,0){5}}

\put(105,20){\line(1,0){5}}
\put(105,21){\line(1,0){5}}
\put(110,18){\mb(4,6){\small $j$}}

\ep
\label{eq:standardzzmeas}
\eea
We can apply \eq{standardzzmeas} to \eq{xtcz4}, and identify the
operations involving the second qubit (from the top) in \eq{xtcz4} as
a two-qubit measurement on the first and third qubits.  This gives a 
simulation circuit for $\Lambda(Z)$: 
\bea
\setlength{\unitlength}{0.6mm}
\bp(80,45)
\put(-17,36){\mb(42,8){\scriptsize 
$X^{\ms a_{\ms 1}} \ms Z^{\ms b_{\ms 1}} \! \ot \!  
 X^{\ms a_{\ms 2}} \ms Z^{\ms b_{\ms 2}} |\psi\rangle$}}

\put(-15,37){\line(0,-1){37}}
\put(-15,0){\vector(1,0){8}}
\put(-13,37){\vector(1,-1){6}}

\put(-3,0){\line(1,0){62}}

\put(31,0){\line(0,1){15}}
\put(29,13){\mb(4,4){$\times$}}
\put(29,-2){\mb(4,4){$\times$}}

\put(34,11){\fb(6,8){$H$}}
\put(40,15){\line(1,0){2}}

\put(42,11){\framebox(10,8){$/$}}
\qbezier(42,11)(47,18)(52,11)
\put(52,15.5){\line(1,0){2.5}}
\put(52,14.5){\line(1,0){2.5}}
\put(53,11){\mb(10,8){$d_{\ms 2}$}}
%---------------------------------------------------------------
\put(26,15){\line(1,0){8}}
\put(6,15){\line(1,0){3}}
\put(-3,15){\line(1,0){3}}
\put(-13,10){\makebox(10,10){$|0\>$}}
\put(0,11){\fb(6,8){$H$}}
%---------------------------------------------------------------
\put(28,-3){\dashbox(33,24){}}

\put(-3,28){\line(1,0){12}}
\put(26,28){\line(1,0){38}}
\put(10,13.5){\fb(15,20){$M_{\!Z \ms \ot \! Z}$}}
\put(9,12.5){\fb(17,22){}}
\put(26,32){\line(1,0){5}}
\put(26,31){\line(1,0){5}}
\put(31,27){\mb(8,10){\small $d_{\ms 1}$}}

%-------------------------------------------------------------

\put(40,36){\mb(60,10){\scriptsize
$ (X^{\ms a_{\ms 1}} \ms Z^{\ms b_{\ms 1}'} \! \ot \! 
 X^{\ms a_{\ms 2}} \ms Z^{\ms b_{\ms 2}'} ) 
\Lambda(Z) |\psi\rangle$}}

\put(75,37){\line(0,-1){37}}
\put(75,0){\vector(-1,0){8}}
\put(73,37){\vector(-1,-1){6}}

\ep
\label{eq:xtcz5} 
\eea
The operations in the dashed box can be implemented by $M_{X \ot Z}$
followed by $M_Z$ on the first qubit (see \eq{intrigue2} in
\sec{circuitmodel}).
With this argument we have rederived Gottesman's remote
controlled-{\sc not} using a single-qubit ancilla and two two-qubit
measurements~\cite{Gottesman98d}, and shown that it is easily
understood as a consequence of one-bit teleportation and the simple
circuit identities \eq{intrigue2} and \eq{zzmeas}.

%%%%%%%%%%%%%%%%%%%%%%%%%%%%%%%%%%%%%%%%%%%%%%%%%%%%%%%%%%%%%%%%%%%%%%
\section{Measurement-based universal quantum computation schemes}

In this section we derive simple variants of both the TQC and 1WQC
models of computation using the principles described in earlier
sections.  Following the discussion in \sec{circuitsim}, it suffices
to show how to perform a universal set of gates in each
measurement-based model of quantum computation.  We will first see
that the simulation circuits derived in the previous section
immediately give a universal scheme in the TQC model.  This scheme is
much simpler than those based on teleportation.  (A similar simplified
scheme was reported independently in \cite{Perdrix04}.)
Then we discuss a method to further reduce the required resources in
the TQC model by identifying and simulating certain subunits of a
circuit.  We then turn to the 1WQC model and present a systematic
derivation of universal quantum computation schemes using primitives
discussed in the previous section.

%%%%%%%%%%%%%%%%%%%%%%%%%%%%%%%%%%%%%%%%%%%%%%%%%%%%%%%%%%%%%%%%%%%%%%
\subsection{Derivation of simplified TQC schemes}
\label{sec:newtqc} 

\subsubsection{Universality}

Consider the universal set consisting of the single-qubit gates and
$\Lambda(Z)$.  A single-qubit gate can be performed in the TQC model using
either \eq{uzt} or \eq{uxt}---the operations in the dashed boxes are
of the form of \eq{intrigue}, with $V$ in the Clifford group and
$V^\dg Z V$ in the Pauli group.  Thus \eq{intrigue2} without the Pauli
correction $V^\dg Z V$ can be used to implement the dashed boxes in
the TQC model.  More concretely, \eq{uzt} and \eq{uxt} imply the
following simulation circuits:
\bea
\setlength{\unitlength}{0.6mm}
\bp(100,25)
\put(-13,10){\mb(20,10){$X^a Z^b |\psi\rangle$}}
\put(0,0){\makebox(10,10){$|0\>$}}

\put(10,13){\line(1,0){4}}
\put(10,5){\line(1,0){4}}

\put(15,3){\fb(30,18){\footnotesize $M_{(U'^\dg X U) \ot X}$}}
\put(14,2){\fb(32,20){}}

\put(46,13){\line(1,0){4}}
\put(46,5){\line(1,0){32}}

\put(46,20.5){\line(1,0){3.5}}
\put(46,19.5){\line(1,0){3.5}}
\put(49,18){\mb(6,4){\footnotesize $c$}}

\put(50,7.5){\fb(24,9){\footnotesize $M_{(U'^\dg Z U)}$}}
\put(51,8.5){\fb(22,7){\footnotesize $M_{(U'^\dg Z U)}$}}

\put(78,0){\makebox(30,10){$X^k Z^c U |\psi\rangle$}}

\put(74,12.5){\line(1,0){4}}
\put(74,13.5){\line(1,0){4}}
\put(78,8){\mb(6,10){$k$}}

\ep
\label{eq:uzttqc}
\\
\setlength{\unitlength}{0.6mm}
\bp(100,25)
\put(-13,10){\mb(20,10){$X^a Z^b |\psi\rangle$}}
\put(0,0){\makebox(10,10){$|+\>$}}

\put(10,13){\line(1,0){4}}
\put(10,5){\line(1,0){4}}

\put(15,3){\fb(30,18){\footnotesize $M_{(U'^\dg Z U) \ot Z}$}}
\put(14,2){\fb(32,20){}}

\put(46,13){\line(1,0){4}}
\put(46,5){\line(1,0){32}}

\put(46,20.5){\line(1,0){3.5}}
\put(46,19.5){\line(1,0){3.5}}
\put(49,18.4){\mb(6,4){\footnotesize $d$}}

\put(50,7.5){\fb(24,9){\footnotesize $M_{(U'^\dg X U)}$}}
\put(51,8.5){\fb(22,7){\footnotesize $M_{(U'^\dg X U)}$}}

\put(78,0){\makebox(30,10){$Z^k X^d U |\psi\rangle$}}

\put(74,12.5){\line(1,0){4}}
\put(74,13.5){\line(1,0){4}}
\put(78,8){\mb(6,10){$k$}}

\ep
\label{eq:uxttqc} 
\eea
In \eqs{uzttqc}{uxttqc} the ancillas can be prepared up to known Pauli
errors that {\em commute} with the subsequent two-qubit measurements.
Simulation circuits for $\Lambda(Z)$ can be obtained from 
\eq{xtcz4} and \eq{xtcz5}:
\bea
\setlength{\unitlength}{0.6mm}
\bp(80,55)
\put(-12,46){\mb(42,8){\scriptsize 
$X^{\ms a_{\ms 1}} \ms Z^{\ms b_{\ms 1}} \! \ot \!  
 X^{\ms a_{\ms 2}} \ms Z^{\ms b_{\ms 2}} |\psi\rangle$}}

\put(-10,47){\line(0,-1){44}}
\put(-10,3){\vector(1,0){8}}
\put(-7,46){\vector(1,-1){7}}

%---------------------------------------------------------------
\put(25,-2){\fb(16,19){\footnotesize$M_{\!X \! \ot \! Z}\!$}}
\put(26,-1){\fb(14,17){}}

\put(41,0){\line(1,0){4}}
\put(41,-1){\line(1,0){4}}
\put(45,-3){\mb(6,4){\footnotesize$d_{\ms 2}$}}

\put(2,3){\line(1,0){23}}
\put(41,3){\line(1,0){14}}

\put(41,15){\line(1,0){3}}

\put(44,9){\fb(10,8){\footnotesize$M_{\!Z}$}}
\put(45,10){\fb(8,6){}}

\put(54,15.5){\line(1,0){3}}
\put(54,14.5){\line(1,0){3}}
\put(56,11){\mb(10,7){\footnotesize $k_2$}}
%---------------------------------------------------------------
\put(11,15){\line(1,0){14}}
\put(2,15){\line(1,0){3}}
\put(-8,10){\makebox(10,10){$|0\>$}}
\put(5,21){\fb(6,8){$H$}}
%---------------------------------------------------------------
\put(5,11){\fb(6,8){$H$}}
\put(-8,20){\makebox(10,10){$|0\>$}}
\put(2,25){\line(1,0){3}}
\put(11,25){\line(1,0){14}}
%---------------------------------------------------------------

\put(41,40){\line(1,0){4}}
\put(41,41){\line(1,0){4}}
\put(45,38.7){\mb(6,4){\footnotesize$d_{\ms 1}$}}

\put(25,23){\fb(16,19){\footnotesize$M_{\!Z \! \ot \! X}\!$}}
\put(26,24){\fb(14,17){}}

\put(44,23){\fb(10,8){\footnotesize$M_{\!Z}$}}
\put(45,24){\fb(8,6){}}

\put(2,38){\line(1,0){23}}
\put(41,38){\line(1,0){14}}

\put(41,25){\line(1,0){3}}

\put(54,25.5){\line(1,0){3}}
\put(54,24.5){\line(1,0){3}}
\put(56,21){\mb(10,7){\footnotesize $k_1$}}

%-------------------------------------------------------------
\put(17,25){\line(0,-1){10}}
\put(15,23){\mb(4,4){$\times$}}
\put(15,13){\mb(4,4){$\times$}}

\put(40,45){\mb(60,10){\scriptsize
$ (X^{\ms a_{\ms 1}} \ms Z^{\ms b_{\ms 1}'} \! \ot \! 
 X^{\ms a_{\ms 2}} \ms Z^{\ms b_{\ms 2}'} ) 
\Lambda(Z) |\psi\rangle$}}

\put(66,47){\line(0,-1){44}}
\put(66,3){\vector(-1,0){8}}
\put(64,46){\vector(-1,-1){7}}

\put(-8,10){\dashbox(28,20){}}
\ep
\label{eq:xtcz4tqc} 
\eea
\bea
\setlength{\unitlength}{0.6mm}
\bp(80,40)
\put(-15,36){\mb(42,8){\scriptsize 
$X^{\ms a_{\ms 1}} \ms Z^{\ms b_{\ms 1}} \! \ot \!  
 X^{\ms a_{\ms 2}} \ms Z^{\ms b_{\ms 2}} |\psi\rangle$}}

\put(-10,37){\line(0,-1){34}}
\put(-10,3){\vector(1,0){8}}
\put(-8,37){\vector(1,-1){8}}

%---------------------------------------------------------------
\put(25,-2){\fb(16,19){\footnotesize$M_{\!X \! \ot \! Z}\!$}}
\put(26,-1){\fb(14,17){}}

\put(41,-1){\line(1,0){4}}
\put(41,0){\line(1,0){4}}
\put(45,-2.5){\mb(6,4){\footnotesize$d_{\ms 2}$}}

\put(1,3){\line(1,0){24}}
\put(41,3){\line(1,0){13}}

\put(41,15){\line(1,0){3}}

\put(44,9){\fb(10,8){\footnotesize$M_{\!Z}$}}
\put(45,10){\fb(8,6){}}

\put(54,15.5){\line(1,0){3}}
\put(54,14.5){\line(1,0){3}}
\put(56,11){\mb(10,7){\footnotesize $k_2$}}
%---------------------------------------------------------------
\put(19,15){\line(1,0){6}}
\put(1,15){\line(1,0){3}}
\put(-8,10){\makebox(8,10){$|+\>$}}
%---------------------------------------------------------------

\put(19,31){\line(1,0){4}}
\put(19,30){\line(1,0){4}}
\put(23,28.5){\mb(6,4){\footnotesize$d_{\ms 1}$}}

\put(1,27){\line(1,0){3}}
\put(19,27){\line(1,0){35}}

\put(5,13.5){\fb(13,18){\small $M_{\!Z\!\ot\!Z}$}}
\put(4,12.5){\fb(15,20){}}

%-------------------------------------------------------------

\put(30,36){\mb(60,10){\scriptsize
$ (X^{\ms a_{\ms 1}} \ms Z^{\ms b_{\ms 1}'} \! \ot \! 
 X^{\ms a_{\ms 2}} \ms Z^{\ms b_{\ms 2}'} ) 
\Lambda(Z) |\psi\rangle$}}

\put(66,37){\line(0,-1){34}}
\put(66,3){\vector(-1,0){8}}
\put(64,37){\vector(-1,-1){8}}

\ep
\label{eq:xtcz5tqc} 
\eea
In the above, $k_1$ should be added to the value of $b_1'$ from
\eq{xtcz4}, and $k_2$ should be added to $b_2'$.
The state $\Lambda(Z) |+\>|+\>$ in \eq{xtcz4tqc} can be prepared by a
two-qubit measurement.
In both \eqs{xtcz4tqc}{xtcz5tqc}, the ancillas can be prepared up to
known $Z$ errors, which have no effect other than flipping the
measurement outcomes of subsequent $M_{\ms X \ms \ot \ms Z \ms}$ and
$M_{\ms Z \ms \ot \ms X \ms}$.
The simulation \eq{xtcz4tqc} uses two ancillary qubits, three
two-qubit measurements, and two single-qubit measurements, and its
logical depth is $3$.  The simulation \eq{xtcz5tqc} uses one ancillary
qubit, two two-qubit measurements, and two single-qubit measurements,
but its logical depth is $4$.

%--------------------------------------------------------------------
\subsubsection{Reduced-cost combined pseudo-simulations}

In the TQC model, how many single- and two-qubit measurements are
required to simulate a circuit $\C$ consisting of single-qubit gates, $m$
$\Lambda(Z)$ gates, and $n$ final single-qubit measurements on the $n$
computation qubits?  We can do better than the method described above
by combining some of the gates in the circuit into larger subunits,
and simulating the subunits directly in the TQC model.
In particular, without loss of generality, there are 
single-qubit gates $U_i, V_i$ for $i = 1,\cdots,m$, such that
$\C$ only consists of $m$ ``composite'' gates
$({U_i}^{\dg}{\ot}{V_i}^{\dg}) \Lambda(Z) (U_i{\ot}V_i)$ applied in order, 
followed by single-qubit measurements. 

Starting from \eq{xtcz4} and \eq{xtcz5} and using \eq{intrigue2},
analogues of \eq{xtcz4tqc} and \eq{xtcz5tqc} can be used to attempt
the simulation of $W = (U^\dg \ot V^\dg) \Lambda(Z) (U \ot V)$ for any
single-qubit gates $U$ and $V$.  These analogues of \eq{xtcz4tqc} and
\eq{xtcz5tqc} simply have $M_{{U'^{\dg}ZU'}\ot{X}}$ and
$M_{{X}{\ot}{V'^{\dg}ZV'}}$ in place of $M_{Z{\ot}X}$ and
$M_{X{\ot}Z}$ respectively.  We call these analogues
``pseudo-simulations,'' because $W$ is simulated up to possible left
multiplications of $U^\dg Z U$ and $V^\dg Z V$, which can easily be
compensated for in the next pseudo-simulation involving the same
qubit.

The complexity of the resulting measurements is comparable to those
required in \eq{uzttqc} and \eq{uxttqc}.  Altogether, a computation
using $m$ $\Lambda(Z)$ gates and $n$ computation qubits can be
pseudo-simulated in TQC using $m$ ancillary qubits, $2m$ two-qubit
measurements, and $2m+n$ single-qubit measurements.  In comparison, a
full simulation (say, using \eq{uzttqc} and \eq{xtcz5tqc}) requires
$3m$ ancillary qubits, $4m$ two-qubit measurements, and $6m+n$
single-qubit measurements.

%%%%%%%%%%%%%%%%%%%%%%%%%%%%%%%%%%%%%%%%%%%%%%%%%%%%%%%%%%%%%%%%%%%%%%
\subsection{Derivation of schemes similar to the 1WQC starting from
the TQC}

In this and subsequent subsections, we present our derivation of
1WQC-like schemes using one-bit teleportation as the underlying
principle, preserving the conceptual simplicity of the TQC.  The
derivation is motivated by the differences between the TQC and 1WQC
models.  The TQC and 1WQC models differ in three main respects:
\begin{enumerate}
\item
The TQC model is similar to the circuit model in that no action is
required on a qubit unless a non-identity gate is applied.  In
contrast, in the 1WQC model, it is necessary to keep measuring qubits
simply to propagate quantum information forward on the lattice.
\item 
  In the TQC model, interactions are effected by multi-qubit
  measurements.  In contrast, no interactions are used in the 1WQC
  model after the initial preparation of the cluster state.  In some
  sense, all interactions are built into the initial state before the
  computation begins.
\item 
  In the 1WQC model, a circuit $\C$ can be simulated using a
  $\C$-dependent graph state, which can in turn be produced from a
  $\C$-independent cluster state.  Thus, the built-in interactions in
  the 1WQC model can be made independent of $\C$.  In contrast, a TQC
  simulation has a one-to-one correspondence with $\C$.
\end{enumerate}

These differences suggest a strategy to derive 1WQC-like schemes using
the principles of the TQC model: every gate is performed by simulation
circuits based on teleportation or one-bit teleportation (such as
\eq{teleportu}, \eq{teleportgc}, \eq{zrot}, \eq{xrot}, and
\eq{xtcz4}).  Suppose the goal is to simulate a circuit $\C$ with $n$
qubits and $m$ time steps.
\begin{enumerate}
\item
  Each gate in $\C$ will be simulated by circuits like \eq{zrot},
  \eq{xrot}, and \eq{xtcz4}.  Furthermore, in each time step, identity
  gates will be explicitly simulated on qubits that are not being
  acted on.  Thus, each qubit will be ``teleported'' in each step.
  Matching the output of one gate simulation to the input of the next,
  we obtain a circuit $\C'$ that ``teleports'' each of the $n$ qubits
  $m$ times, with the desired gates performed along the way.  $\C'$
  contains initial $|0\>$ states, one- or two-qubit gates, and
  single-qubit measurements.  Note that in this circuit we do not
  interpret a two-qubit gate followed by a single-qubit measurement as an
  incomplete two-qubit measurement, as we did in the TQC.  The reason is
  that in the next step we will build the two-qubit gates into the
  initial state, leading to an equivalent circuit containing only
  single-qubit measurements.
\item 
  To build interactions into the initial state, we apply standard
  circuit identities to rewrite $\C'$ so that all two-qubit gates
  occur before the $\C$-dependent single-qubit gates, followed by
  single-qubit measurements.
  The circuits used to simulate each gate are chosen to facilitate
  this step.
  The resulting circuit $\C''$ consists of ($i$) two-qubit gates
  acting on circuit-independent product states, ($ii$)
  circuit-dependent single-qubit gates followed by single-qubit
  measurements.
  We regard the state $|\psi_\C\>$ after step ($i$) as a new initial
  state, and the remaining single-qubit gates and measurement in step
  ($ii$) as single-qubit measurements in redefined bases.
  We can thus interpret $\C''$ as starting from a $\C$-dependent
  initial state $|\psi_\C\>$, followed by single-qubit measurements.
  We will see that $|\psi_\C\>$ is like the circuit-dependent graph
  state in the original 1WQC scheme.
  Schemes derived in this way will be called 1WQC$_{\rm TG}$ schemes,
  with $T$ standing for the underlying principle of teleportation, and
  $G$ for an initial graph state.
\item 
  We want to modify the 1WQC$_{\rm TG}$ schemes to start with a fixed,
  universal initial state analogous to the cluster state.  The idea
  is to find a circuit that simulates a two-qubit gate or the identity
  gate depending on the choices of the single-qubit measurements.  In
  other words,
  the interactions built into the initial state are ``undoable,'' in
  the sense that they may be optionally removed by some later
  single-qubit measurement.  The desired universal initial state simply
  has an undoable interaction built in wherever the interaction may
  occur.  We call the resulting model 1WQC$_{\rm T}$.
\end{enumerate}

%%%%%%%%%%%%%%%%%%%%%%%%%%%%%%%%%%%%%%%%%%%%%%%%%%%%%%%%%%%%%%%%%%%%%%%%%
\subsection{Derivation of schemes starting from a circuit-dependent
graph state}
\label{sec:1wqctg}

%------------------------------------------------------------------------
\subsubsection{A universal circuit decomposition}
\label{sec:universaldecomp}

The most general quantum circuit $\C$ can be viewed as consisting of
alternating steps of ($i$) arbitrary single-qubit gates and ($ii$)
optional {\em nearest-neighbor} $\Lambda(Z)$ gates (because $H$ and
$\Lambda(Z)$ can be composed to make swap gates).
We want gate-simulation circuits in which the interactions can be
performed before the $\C$-dependent single-qubit gates.  Thus, simulation
circuits like \eq{zrot} and \eq{xrot} are preferred to ones like \eq{uzt} and
\eq{uxt}.
Such choices preserve universality since any single-qubit gate has an
Euler angle decomposition $U = Z_{\t_3} X_{\t_2} Z_{\t_1}$.
The circuit $\C$ now contains cycles of ($i$) arbitrary $Z$ rotations,
($ii$) arbitrary $X$ rotations, ($iii$) arbitrary $Z$ rotations, and
($iv$) optional nearest-neighbor $\Lambda(Z)$ gates, i.e., $\Lambda(Z)^k$ where
$k$ can be freely chosen from $\{0,1\}$.
Since a $\Lambda(Z)$ commutes with the $Z$ rotations before and after, the
$Z$ rotations can be merged.  For example, two cycles on two qubits
can be represented by
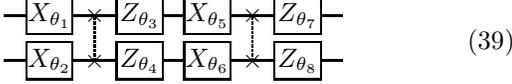
\bea
\setlength{\unitlength}{0.6mm}
\centering
\bp(95,20)
\put(10,5){\line(1,0){5}}
\put(10,15){\line(1,0){5}}
\put(15,1){\fb(10,8){$X_{\t_2}$}}
\put(15,11){\fb(10,8){$X_{\t_1}$}}
\put(25,5){\line(1,0){10}}
\put(25,15){\line(1,0){10}}

\multiput(30,5)(0,1){10}{\line(0,1){0.5}}
\put(28,3){\mb(4,4){$\times$}}
\put(28,13){\mb(4,4){$\times$}}

\put(35,1){\fb(10,8){$Z_{\t_4}$}}
\put(35,11){\fb(10,8){$Z_{\t_3}$}}
\put(45,5){\line(1,0){5}}
\put(45,15){\line(1,0){5}}

\put(50,1){\fb(10,8){$X_{\t_6}$}}
\put(50,11){\fb(10,8){$X_{\t_5}$}}
\put(60,5){\line(1,0){10}}
\put(60,15){\line(1,0){10}}

\multiput(65,5)(0,1){10}{\line(0,1){0.5}}
\put(63,3){\mb(4,4){$\times$}}
\put(63,13){\mb(4,4){$\times$}}

\put(70,1){\fb(10,8){$Z_{\t_8}$}}
\put(70,11){\fb(10,8){$Z_{\t_7}$}}
\put(80,5){\line(1,0){5}}
\put(80,15){\line(1,0){5}}
\ep
\label{eq:circuiteg} 
\eea
where $\t_i$ are arbitrary angles of rotation, and the dotted line for
$\Lambda(Z)$ denotes an optional gate.
We will see that it is more efficient to simulate 
$\Lambda(Z)$ and $Z$ rotations together.  
Thus, a circuit should be decomposed into cycles, each with two steps:
($i$) arbitrary $X$ rotations, and ($ii$) arbitrary $Z$ rotations and
optional nearest-neighbor $\Lambda(Z)$ gates.
 
%%%%%%%%%%%%%%%%%%%%%%%%%%%%%%%%%%%%%%%%%%%%%%%%%%%%%%%%%%%%%%%%%%%%%%%%%
\subsubsection{Simulation using one-bit teleportation}
\label{sec:1bit-telep}

We first describe the simulation circuits for the elementary steps
just described.  We use \eq{xrot} to simulate $X_\t$, restated here:
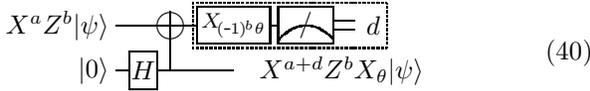
\bea
\setlength{\unitlength}{0.6mm}
\centering
\bp(80,20)
\put(-14,10){\mb(22,10){$X^a Z^b|\psi\rangle$}}
\put(0,0){\makebox(10,10){$|0\>$}}

\put(10,15){\line(1,0){18}}
\put(10,5){\line(1,0){3}}

\put(22,18){\line(0,-1){13}}
\put(22,15){\circle{6}}

\put(13,1){\fb(6,8){$H$}}
\put(19,5){\line(1,0){17}}

\put(28,11){\fb(16,8){\footnotesize $X_{\! (\mbox{-} \ms 1 \ms )^{\ms
\ms  b} \t}$}}
\put(44,15){\line(1,0){2}}

\put(27,10){\dashbox{0.5}(43,10){}}

\put(35, 0){\makebox(50,10){$X^{a+d} Z^{b} X_\t |\psi\rangle$}}

\put(46,11){\framebox(12,8){$/$}}
\qbezier(46,12)(52,18)(58,12)
\put(58,16){\line(1,0){5}}
\put(58,14){\line(1,0){5}}
\put(62,10){\mb(10,10){$d$}}

\ep
\label{eq:xrot2} 
\eea
We will identify $H|0\> = |+\>$ as part of the initial state
preparation.
We simulate an optional $\Lambda(Z)$ gate and $Z$ rotations in a single step
as follows:
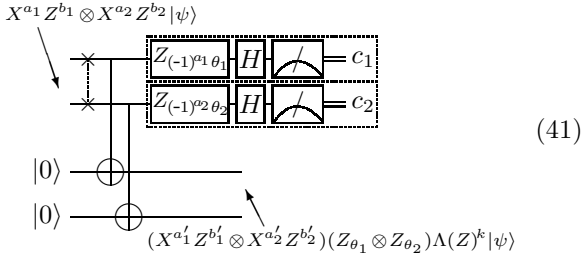
\bea
\setlength{\unitlength}{0.6mm}
\bp(80,50)
\put(-12,46){\mb(42,8){\scriptsize 
$X^{\ms a_{\ms 1}} \ms Z^{\ms b_{\ms 1}} \! \ot \! 
 X^{\ms a_{\ms 2}} \ms Z^{\ms b_{\ms 2}} |\psi\rangle$}}
\put(-8,0){\makebox(10,10){$|0\>$}}
\put(-5,47){\vector(1,-3){4.5}}

\put(2,30){\line(1,0){18}}
\put(2,5){\line(1,0){38}}

\put(15,30){\line(0,-1){28}}
\put(15,5){\circle{6}}

\put(20,26){\fb(17,8){\footnotesize 
$Z_{\ms (\ms \mbox{-} \ms 1 \ms )^{\!a_{\!2}} \ms \t_{\ms 2}} \!$}}
\put(37,30){\line(1,0){2}}

\put(39,26){\fb(6,8){$H$}}
\put(45,30){\line(1,0){2}}

\put(19,25){\dashbox{0.5}(51,10){}}

\put(47,26){\framebox(11,8){$/$}}
\qbezier(47,26)(52,33)(58,26)
\put(58,30.5){\line(1,0){5}}
\put(58,29.5){\line(1,0){5}}
\put(62,26){\mb(10,8){$c_2$}}

%---------------------------------------------------------------
\put(-8,10){\makebox(10,10){$|0\>$}}

\put(2,40){\line(1,0){18}}
\put(2,15){\line(1,0){38}}

\put(11,40){\line(0,-1){28}}
\put(11,15){\circle{6}}

\put(20,36){\fb(17,8){\footnotesize 
$Z_{\ms (\ms \mbox{-} \ms 1 \ms)^{\! a_{\!1}} \ms \t_{\ms 1}} \!$}}
\put(37,40){\line(1,0){2}}

\put(39,36){\fb(6,8){$H$}}
\put(45,40){\line(1,0){2}}

\put(19,35){\dashbox{0.5}(51,10){}}

\put(47,36){\framebox(11,8){$/$}}
\qbezier(47,36)(52,43)(58,36)
\put(58,40.5){\line(1,0){5}}
\put(58,39.5){\line(1,0){5}}
\put(62,36){\mb(10,8){$c_1$}}

%-------------------------------------------------------------
\multiput(6,40)(0,-1){10}{\line(0,-1){0.5}}
\put(4,38){\mb(4,4){$\times$}}
\put(4,28){\mb(4,4){$\times$}}

\put(20,-5){\mb(80,10){\scriptsize
$ (X^{\ms a_{\ms 1}'} \ms Z^{\ms b_{\ms 1}'} \! \ot \! 
 X^{\ms a_{\ms 2}'} \ms Z^{\ms b_{\ms 2}'} ) 
(Z_{\t_1} \! \ot \! Z_{\t_2}) \Lambda(Z)^{\!k} |\psi\rangle$}}

\put(45,3){\vector(-1,2){4}}

\ep
\label{eq:1bittelepcz} 
\eea
In the above, $\Lambda(Z)$ is performed if $k=1$, and not if $k=0$.  The
state after $\Lambda(Z)^k$ is 
$ (X^{\ms a_{\ms 1}} \ms Z^{\ms b_{\ms 1} + a_{\ms 2} k} \! \ot \!
 X^{\ms a_{\ms 2}} \ms Z^{\ms b_{\ms 2} + a_{\ms 1} k } ) \, \Lambda(Z)^{k}
 |\psi\rangle$.  
After the $Z$ rotations and teleportation, the final output is
$ (X^{\ms a_{\ms 1}} \ms Z^{\ms b_{\ms 1} + a_{\ms 2} k + c_1} \! \ot \!
 X^{\ms a_{\ms 2}} \ms Z^{\ms b_{\ms 2} + a_{\ms 1} k + c_2 } ) \,(Z_{\t_1}
 \! \ot \! Z_{\t_2}) \Lambda(Z)^{k} |\psi\rangle$.
Therefore,
$a_1' = a_1$, 
$a_2' = a_2$, 
$b_1' = b_{\ms 1} + a_{\ms 2} k + c_1$, and
$b_2' = b_{\ms 2} + a_{\ms 1} k + c_2$ 
in \eq{1bittelepcz}.

Finally, we chain together the simulation circuits for the repeating
cycles of ($i$) arbitrary $X$ rotations and ($ii$) arbitrary $Z$
rotations and optional nearest-neighbor $\Lambda(Z)$ gates.
The resulting circuit to
simulate \eq{circuiteg}, with two cycles for two qubits, is
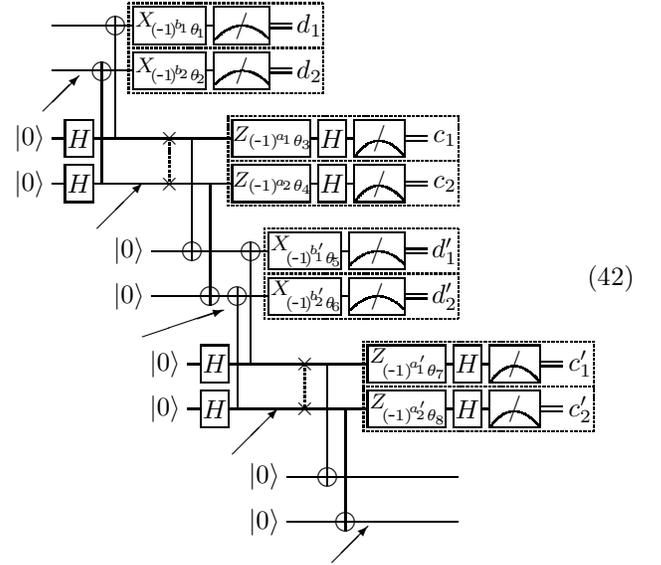
\bea
\setlength{\unitlength}{0.6mm}
\hspace*{25ex}
\bp(70,120)

\put(-60,75){\makebox(10,10){$|0\>$}}

\put(-50,105){\line(1,0){18}}
\put(-50,80){\line(1,0){3}}

\put(-39,107){\line(0,-1){27}}
\put(-39,105){\circle{4}}

\put(-47,76){\fb(6,8){$H$}}
\put(-41,80){\line(1,0){17}}

\put(-32,101){\fb(16,8){\footnotesize 
$X_{\!(\ms \mbox{-} \ms 1\!)^{\!b_{\!2}} \ms \t_{\!2}}\!$}}
\put(-16,105){\line(1,0){2}}

\put(-33,100){\dashbox{0.5}(43,10){}}

\put(-14,101){\framebox(12,8){$/$}}
\qbezier(-14,102)(-8,108)(-2,102)
\put(-2,105.5){\line(1,0){5}}
\put(-2,104.5){\line(1,0){5}}
\put(2,101){\mb(10,8){$d_2$}}

%-----------------------------------------------------------------------

\put(-60,85){\makebox(10,10){$|0\>$}}

\put(-50,115){\line(1,0){18}}
\put(-50,90){\line(1,0){3}}

\put(-36,117){\line(0,-1){27}}
\put(-36,115){\circle{4}}

\put(-47,86){\fb(6,8){$H$}}
\put(-41,90){\line(1,0){17}}

\put(-32,111){\fb(16,8){\footnotesize 
$X_{\!(\ms \mbox{-} \ms 1\!)^{\!b_{\!1}} \ms \t_{\!1}}\!$}}

\put(-16,115){\line(1,0){2}}

\put(-33,110){\dashbox{0.5}(43,10){}}

\put(-14,111){\framebox(12,8){$/$}}
\qbezier(-14,112)(-8,118)(-2,112)
\put(-2,115.5){\line(1,0){5}}
\put(-2,114.5){\line(1,0){5}}
\put(2,111){\mb(10,8){$d_1$}}

%-----------------------------------------------------------------------
\put(-38,50){\makebox(10,10){$|0\>$}}

\put(-28,80){\line(1,0){18}}

\put(-28,55){\line(1,0){18}}

\put(-15,80){\line(0,-1){27}}
\put(-15,55){\circle{4}}

\put(-10,76){\fb(17,8){\footnotesize 
$Z_{(\ms \mbox{-} \ms 1)^{\!a_{\!2}} \ms \t_{\ms 4}}\!$}}
\put(7,80){\line(1,0){2}}

\put(9,76){\fb(6,8){$H$}}
\put(15,80){\line(1,0){2}}

\put(-11,75){\dashbox{0.5}(51,10){}}

\put(17,76){\framebox(11,8){$/$}}
\qbezier(17,76)(22,83)(28,76)
\put(28,80.5){\line(1,0){5}}
\put(28,79.5){\line(1,0){5}}
\put(32,76){\mb(10,8){$c_2$}}

%---------------------------------------------------------------
\put(-38,60){\makebox(10,10){$|0\>$}}

\put(-28,90){\line(1,0){18}}
\put(-28,65){\line(1,0){18}}

\put(-19,90){\line(0,-1){27}}
\put(-19,65){\circle{4}}

\put(-10,86){\fb(17,8){\footnotesize 
$Z_{(\ms \mbox{-} \ms 1)^{\!a_{\!1}} \ms \t_{\ms 3}}\!$}}
\put(7,90){\line(1,0){2}}

\put( 9,86){\fb(6,8){$H$}}
\put(15,90){\line(1,0){2}}

\put(-11,85){\dashbox{0.5}(51,10){}}

\put(17,86){\framebox(11,8){$/$}}
\qbezier(17,86)(22,93)(28,86)
\put(28,90.5){\line(1,0){5}}
\put(28,89.5){\line(1,0){5}}
\put(32,86){\mb(10,8){$c_1$}}

%-------------------------------------------------------------
\multiput(-24,90)(0,-1){10}{\line(0,-1){0.5}}
\put(-26,88){\mb(4,4){$\times$}}
\put(-26,78){\mb(4,4){$\times$}}

%-------------------------------------------------------------
%-------------------------------------------------------------
\put(-30,25){\makebox(10,10){$|0\>$}}

\put(-20,55){\line(1,0){18}}
\put(-20,30){\line(1,0){3}}

\put(-9,57){\line(0,-1){27}}
\put(-9,55){\circle{4}}

\put(-17,26){\fb(6,8){$H$}}
\put(-11,30){\line(1,0){17}}

\put(-2,51){\fb(16,8){\footnotesize 
$X_{\!(\ms \mbox{-} \ms 1\!)^{\!b_{\!2}'} \ms \t_{\!6}}\!$}}
\put(14,55){\line(1,0){2}}

\put(-3,50){\dashbox{0.5}(43,10){}}

\put(16,51){\framebox(12,8){$/$}}
\qbezier(16,52)(22,58)(28,52)
\put(28,55.5){\line(1,0){5}}
\put(28,54.5){\line(1,0){5}}
\put(32,51){\mb(10,8){$d_2'$}}

%-----------------------------------------------------------------------

\put(-30,35){\makebox(10,10){$|0\>$}}

\put(-20,65){\line(1,0){18}}
\put(-20,40){\line(1,0){3}}

\put(-6,67){\line(0,-1){27}}
\put(-6,65){\circle{4}}

\put(-17,36){\fb(6,8){$H$}}
\put(-11,40){\line(1,0){17}}

\put(-2,61){\fb(16,8){\footnotesize 
$X_{\!(\ms \mbox{-} \ms 1\!)^{\!b_{\!1}'} \ms \t_{\!5}}\!$}}
\put(14,65){\line(1,0){2}}

\put(-3,60){\dashbox{0.5}(43,10){}}

\put(16,61){\framebox(12,8){$/$}}
\qbezier(16,62)(22,68)(28,62)
\put(28,65.5){\line(1,0){5}}
\put(28,64.5){\line(1,0){5}}
\put(32,61){\mb(10,8){$d_1'$}}

%-----------------------------------------------------------------------
\put(-8,0){\makebox(10,10){$|0\>$}}

\put(2,30){\line(1,0){18}}
\put(2,5){\line(1,0){38}}

\put(15,30){\line(0,-1){27}}
\put(15,5){\circle{4}}

\put(20,26){\fb(17,8){\footnotesize 
$Z_{\!(\ms \mbox{-} \ms 1)^{\!a_{\!2}'} \ms \t_{\ms 8}}\!$}}
\put(37,30){\line(1,0){2}}

\put(39,26){\fb(6,8){$H$}}
\put(45,30){\line(1,0){2}}

\put(19,25){\dashbox{0.5}(51,10){}}

\put(47,26){\framebox(11,8){$/$}}
\qbezier(47,26)(52,33)(58,26)
\put(58,30.5){\line(1,0){5}}
\put(58,29.5){\line(1,0){5}}
\put(62,26){\mb(10,8){$c_2'$}}

%---------------------------------------------------------------
\put(-8,10){\makebox(10,10){$|0\>$}}

\put(2,40){\line(1,0){18}}
\put(2,15){\line(1,0){38}}

\put(11,40){\line(0,-1){27}}
\put(11,15){\circle{4}}

\put(20,36){\fb(17,8){\footnotesize 
$Z_{\!(\ms \mbox{-} \ms 1)^{\!a_{\!1}'} \ms \t_{\ms 7}}\!$}}
\put(37,40){\line(1,0){2}}

\put(39,36){\fb(6,8){$H$}}
\put(45,40){\line(1,0){2}}

\put(19,35){\dashbox{0.5}(51,10){}}

\put(47,36){\framebox(11,8){$/$}}
\qbezier(47,36)(52,43)(58,36)
\put(58,40.5){\line(1,0){5}}
\put(58,39.5){\line(1,0){5}}
\put(62,36){\mb(10,8){$c_1'$}}

%-------------------------------------------------------------
\multiput(6,40)(0,-1){10}{\line(0,-1){0.5}}
\put(4,38){\mb(4,4){$\times$}}
\put(4,28){\mb(4,4){$\times$}}

%-------------------------------------------------------------

\put(-40,70){\vector(1,1){10}}
\put(-10,20){\vector(1,1){10}}

\put(-30,46){\vector(3,1){18}}
\put(-52,96){\vector(1,1){8}}
\put(12,-4){\vector(1,1){8}}

\ep
\label{eq:1bitteleport-chain} 
\eea
Each arrow in \eq{1bitteleport-chain} indicates where the output of a
certain teleportation step matches the input of the subsequent
teleportation.
The values of $a_i'$ and $b_i'$ can be read from \eq{xrot2} and
\eq{1bittelepcz}.
The circuit of~\eq{1bitteleport-chain} generalizes easily to $n$
qubits with multiple optional $\Lambda(Z)$ gates.

The simulation \eq{1bitteleport-chain} can be simplified by ($i$)
rewriting $\Lambda(X)$ as $(I{\ot}H) \Lambda(Z) (I{\ot}H)$, ($ii$) canceling
out consecutive Hadamard gates (since $H^2 = I$), ($iii$) rewriting
$H|0\>$ as $|+\>$, and ($iv$) absorbing $H$ before a single-qubit
measurement as part of the measurement.  We thus obtain a simpler
simulation scheme for \eq{circuiteg}:
\bea
\setlength{\unitlength}{0.6mm}
\hspace*{20ex}
\bp(70,125)

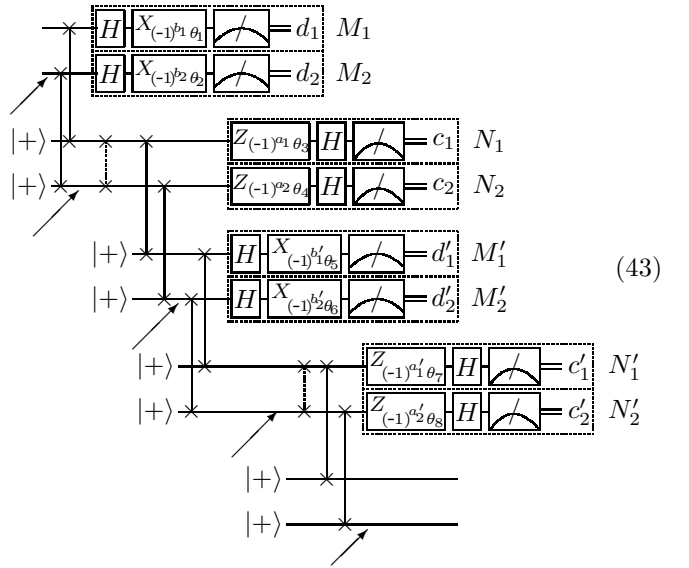
\put(-60,75){\makebox(10,10){$|+\>$}}

\put(-52,105){\line(1,0){12}}
\put(-50,80){\line(1,0){10}}

\put(-40,101){\fb(6,8){$H$}}
\put(-34,105){\line(1,0){2}}
\put(-41,80){\line(1,0){18}}

\put(-48,105){\line(0,-1){25}}
\put(-50,103){\mb(4,4){$\times$}}
\put(-50,78){\mb(4,4){$\times$}}

\put(-41,80){\line(1,0){17}}

\put(-32,101){\fb(16,8){\footnotesize 
$X_{\!(\ms \mbox{-} \ms 1\!)^{\!b_{\!2}} \ms \t_{\!2}}\!$}}
\put(-16,105){\line(1,0){2}}

\put(-41,100){\dashbox{0.5}(51,10){}}
\put(12,100){\mb(10,10){$M_2$}}

\put(-14,101){\framebox(12,8){$/$}}
\qbezier(-14,102)(-8,108)(-2,102)
\put(-2,105.5){\line(1,0){5}}
\put(-2,104.5){\line(1,0){5}}
\put(2,101){\mb(10,8){$d_2$}}

%-----------------------------------------------------------------------

\put(-60,85){\makebox(10,10){$|+\>$}}

\put(-52,115){\line(1,0){12}}
\put(-50,90){\line(1,0){10}}

\put(-46,115){\line(0,-1){25}}
\put(-48,113){\mb(4,4){$\times$}}
\put(-48,88){\mb(4,4){$\times$}}

\put(-41,90){\line(1,0){17}}

\put(-40,111){\fb(6,8){$H$}}
\put(-34,115){\line(1,0){2}}
\put(-41,90){\line(1,0){18}}

\put(-32,111){\fb(16,8){\footnotesize 
$X_{\!(\ms \mbox{-} \ms 1\!)^{\!b_{\!1}} \ms \t_{\!1}}\!$}}

\put(-16,115){\line(1,0){2}}

\put(-41,110){\dashbox{0.5}(51,10){}}
\put(12,110){\mb(10,10){$M_1$}}

\put(-14,111){\framebox(12,8){$/$}}
\qbezier(-14,112)(-8,118)(-2,112)
\put(-2,115.5){\line(1,0){5}}
\put(-2,114.5){\line(1,0){5}}
\put(2,111){\mb(10,8){$d_1$}}

%-----------------------------------------------------------------------
\put(-42,50){\makebox(10,10){$|+\>$}}

\put(-28,80){\line(1,0){18}}

\put(-32,55){\line(1,0){18}}

\put(-25,80){\line(0,-1){25}}
\put(-27,78){\mb(4,4){$\times$}}
\put(-27,53){\mb(4,4){$\times$}}

\put(-10,76){\fb(17,8){\footnotesize 
$Z_{(\ms \mbox{-} \ms 1)^{\!a_{\!2}} \ms \t_{\ms 4}}\!$}}
\put(7,80){\line(1,0){2}}

\put(9,76){\fb(6,8){$H$}}
\put(15,80){\line(1,0){2}}

\put(-11,75){\dashbox{0.5}(51,10){}}
\put(42,75){\mb(10,10){$N_2$}}

\put(17,76){\framebox(11,8){$/$}}
\qbezier(17,76)(22,83)(28,76)
\put(28,80.5){\line(1,0){5}}
\put(28,79.5){\line(1,0){5}}
\put(32,76){\mb(10,8){$c_2$}}

%---------------------------------------------------------------
\put(-42,60){\makebox(10,10){$|+\>$}}

\put(-28,90){\line(1,0){18}}
\put(-32,65){\line(1,0){18}}

\put(-29,90){\line(0,-1){25}}
\put(-31,63){\mb(4,4){$\times$}}
\put(-31,88){\mb(4,4){$\times$}}

\put(-10,86){\fb(17,8){\footnotesize 
$Z_{(\ms \mbox{-} \ms 1)^{\!a_{\!1}} \ms \t_{\ms 3}}\!$}}
\put(7,90){\line(1,0){2}}

\put( 9,86){\fb(6,8){$H$}}
\put(15,90){\line(1,0){2}}

\put(-11,85){\dashbox{0.5}(51,10){}}
\put(42,85){\mb(10,10){$N_1$}}

\put(17,86){\framebox(11,8){$/$}}
\qbezier(17,86)(22,93)(28,86)
\put(28,90.5){\line(1,0){5}}
\put(28,89.5){\line(1,0){5}}
\put(32,86){\mb(10,8){$c_1$}}

%-------------------------------------------------------------
\multiput(-38,90)(0,-1){10}{\line(0,-1){0.5}}
\put(-40,88){\mb(4,4){$\times$}}
\put(-40,78){\mb(4,4){$\times$}}

%-------------------------------------------------------------
%-------------------------------------------------------------
\put(-32,25){\makebox(10,10){$|+\>$}}

\put(-22,55){\line(1,0){12}}
\put(-22,30){\line(1,0){13}}

\put(-19,55){\line(0,-1){25}}
\put(-21,53){\mb(4,4){$\times$}}
\put(-21,28){\mb(4,4){$\times$}}

\put(-10,51){\fb(6,8){$H$}}
\put(-4,55){\line(1,0){2}}
\put(-11,30){\line(1,0){18}}

\put(-2,51){\fb(16,8){\footnotesize 
$X_{\!(\ms \mbox{-} \ms 1\!)^{\!b_{\!2}'} \! \t_{\!6}}\!$}}
\put(14,55){\line(1,0){2}}

\put(-11,50){\dashbox{0.5}(51,10){}}
\put(42,50){\mb(10,10){$M_2'$}}

\put(16,51){\framebox(12,8){$/$}}
\qbezier(16,52)(22,58)(28,52)
\put(28,55.5){\line(1,0){5}}
\put(28,54.5){\line(1,0){5}}
\put(32,51){\mb(10,8){$d_2'$}}

%-----------------------------------------------------------------------

\put(-32,35){\makebox(10,10){$|+\>$}}

\put(-20,65){\line(1,0){10}}

\put(-22,40){\line(1,0){13}}

\put(-16,65){\line(0,-1){25}}
\put(-18,63){\mb(4,4){$\times$}}
\put(-18,38){\mb(4,4){$\times$}}

\put(-10,61){\fb(6,8){$H$}}
\put(-4,65){\line(1,0){2}}

\put(-11,40){\line(1,0){17}}

\put(-2,61){\fb(16,8){\footnotesize 
$X_{\!(\ms \mbox{-} \ms 1\!)^{\!b_{\!1}'} \! \t_{\!5}}\!$}}
\put(14,65){\line(1,0){2}}

\put(-11,60){\dashbox{0.5}(51,10){}}
\put(42,60){\mb(10,10){$M_1'$}}

\put(16,61){\framebox(12,8){$/$}}
\qbezier(16,62)(22,68)(28,62)
\put(28,65.5){\line(1,0){5}}
\put(28,64.5){\line(1,0){5}}
\put(32,61){\mb(10,8){$d_1'$}}

%-----------------------------------------------------------------------
\put(-8,0){\makebox(10,10){$|+\>$}}

\put(2,30){\line(1,0){18}}
\put(2,5){\line(1,0){38}}

\put(15,30){\line(0,-1){25}}
\put(13,28){\mb(4,4){$\times$}}
\put(13,3){\mb(4,4){$\times$}}

\put(20,26){\fb(17,8){\footnotesize 
$Z_{\!(\ms \mbox{-} \ms 1)^{\!a_{\!2}'} \ms \t_{\ms 8}}\!$}}
\put(37,30){\line(1,0){2}}

\put(39,26){\fb(6,8){$H$}}
\put(45,30){\line(1,0){2}}

\put(72,25){\mb(10,10){$N_2'$}}
\put(19,25){\dashbox{0.5}(51,10){}}

\put(47,26){\framebox(11,8){$/$}}
\qbezier(47,26)(52,33)(58,26)
\put(58,30.5){\line(1,0){5}}
\put(58,29.5){\line(1,0){5}}
\put(62,26){\mb(10,8){$c_2'$}}

%---------------------------------------------------------------
\put(-8,10){\makebox(10,10){$|+\>$}}

\put(2,40){\line(1,0){18}}
\put(2,15){\line(1,0){38}}

\put(11,40){\line(0,-1){25}}
\put(9,38){\mb(4,4){$\times$}}
\put(9,13){\mb(4,4){$\times$}}

\put(20,36){\fb(17,8){\footnotesize 
$Z_{\!(\ms \mbox{-} \ms 1)^{\!a_{\!1}'} \ms \t_{\ms 7}}\!$}}
\put(37,40){\line(1,0){2}}

\put(39,36){\fb(6,8){$H$}}
\put(45,40){\line(1,0){2}}

\put(19,35){\dashbox{0.5}(51,10){}}
\put(72,35){\mb(10,10){$N_1'$}}

\put(47,36){\framebox(11,8){$/$}}
\qbezier(47,36)(52,43)(58,36)
\put(58,40.5){\line(1,0){5}}
\put(58,39.5){\line(1,0){5}}
\put(62,36){\mb(10,8){$c_1'$}}

%-------------------------------------------------------------
\multiput(6,40)(0,-1){10}{\line(0,-1){0.5}}
\put(4,38){\mb(4,4){$\times$}}
\put(4,28){\mb(4,4){$\times$}}

%-------------------------------------------------------------

\put(-54,69){\vector(1,1){10}}
\put(-10,20){\vector(1,1){10}}

\put(-32,44){\vector(1,1){10}}
\put(-59,96){\vector(1,1){8}}
\put(12,-4){\vector(1,1){8}}

\ep
\label{eq:1bit-chain}
\eea
We can view the operations in the dashed boxes as single-qubit
measurements, and the rest of the circuit as an initial state
$|\psi_\C\>$.  Note that $|\psi_\C\>$ is a graph state, as defined in
\sec{1wqcmodel}.  The circuit $\C$ determines whether each $\Lambda(Z)^k=I$
or $\Lambda(Z)$ in $|\psi_\C\>$.
The substrate representation of \eq{1bit-chain} is 
\bea
\setlength{\unitlength}{0.6mm}
\centering
\bp(90,25)
\put(10,5){\line(1,0){5}}
\put(10,20){\line(1,0){5}}
\put(15,1){\fb(10,8){$M_2$}}
\put(15,16){\fb(10,8){$M_1$}}
\put(25,5){\line(1,0){7.5}}
\put(25,20){\line(1,0){7.5}}

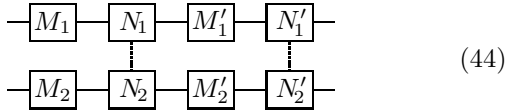
\multiput(37.5,9)(0,1){7}{\line(0,1){0.5}}

\put(32.5,1){\fb(10,8){$\,N_{\!2}$}}
\put(32.5,16){\fb(10,8){$\,N_{\!1}$}}
\put(42.5,5){\line(1,0){7.5}}
\put(42.5,20){\line(1,0){7.5}}

\put(50,1){\fb(10,8){$M_2'$}}
\put(50,16){\fb(10,8){$M_1'$}}
\put(60,5){\line(1,0){7.5}}
\put(60,20){\line(1,0){7.5}}

\multiput(72.5,9)(0,1){7}{\line(0,1){0.5}}

\put(67.5,1){\fb(10,8){$\,N_{\!2}'$}}
\put(67.5,16){\fb(10,8){$\,N_{\!1}'$}}
\put(77.5,5){\line(1,0){5}}
\put(77.5,20){\line(1,0){5}}
\ep
\label{eq:1bitsub} 
\eea
where the measurements are as specified in \eq{1bit-chain}.  The
initial graph state for an $n$-qubit circuit $\C$ with $m$ time steps
can be chosen to be
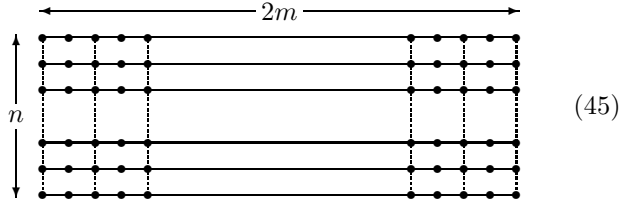
\bea
\setlength{\unitlength}{0.35mm}
\centering
\bp(180,80)

\put(-10,30){\vector(0,-1){26}}
\put(-10,40){\vector(0,1){26}}
\put(-15,30){\mb(10,10){$n$}}

\put(81,75){\vector(-1,0){82}}
\put(99,75){\vector(1,0){82}}
\put(85,70){\mb(10,10){$2m$}}

\multiput(0,5)(0,10){3}{\line(1,0){180}}
\multiput(0,45)(0,10){3}{\line(1,0){180}}

\multiput(0,5)(20,00){3}{\multiput(0,0)(0,2){30}{\line(0,1){1}}}
\multiput(140,5)(20,00){3}{\multiput(0,0)(0,2){30}{\line(0,1){1}}}

\multiput(0,5)(0,10){3}{   \multiput(0,0)(10,00){5}{\circle*{3}}  }
\multiput(0,45)(0,10){3}{   \multiput(0,0)(10,00){5}{\circle*{3}}  }

\multiput(140,5)(0,10){3}{   \multiput(0,0)(10,00){5}{\circle*{3}}  }
\multiput(140,45)(0,10){3}{   \multiput(0,0)(10,00){5}{\circle*{3}}  }

\ep
\label{eq:1wqctg} 
\eea

%%%%%%%%%%%%%%%%%%%%%%%%%%%%%%%%%%%%%%%%%%%%%%%%%%%%%%%%%%%%%%%%%%%%%%%%%
\subsection{Derivation of schemes starting from a universal initial state}
\label{sec:universal}

We now present methods for performing an undoable $\Lambda(Z)$.  Composing
the undoable $\Lambda(Z)$ simulation with the 1WQC$_{\rm TG}$ scheme
described in the previous section, we obtain various
1WQC$_{\rm T}$ schemes.

%%%%%%%%%%%%%%%%%%%%%%%%%%%%%%%%%%%%%%%%%%%%%%%%%%%%%%%%%%%%%%%%%%%%%%%%%
\subsubsection{The remote $\Lambda(Z)$ approach (I)}

Consider the circuit \eq{xtcz4} that simulates $\Lambda(Z)$: 
\bea
\setlength{\unitlength}{0.6mm}
\centering
\bp(100,65)

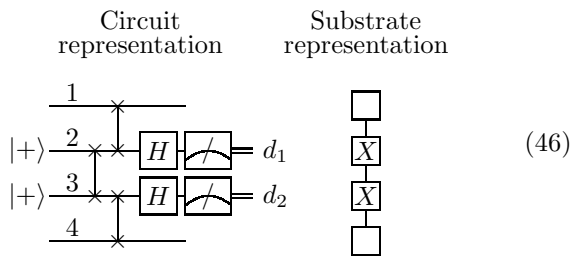
\put(80,33){\line(0,1){4}}
\put(80,23){\line(0,1){4}}
\put(80,13){\line(0,1){4}}

\put(77,37){\fb(6,6){}}
\put(77,27){\fb(6,6){$X$}}
\put(77,17){\fb(6,6){$X$}}
\put(77, 7){\fb(6,6){}}

\put(0,54){\mb(60,10){Circuit}}
\put(0,48){\mb(60,10){representation}}
\put(60,54){\mb(40,10){Substrate}}
\put(60,48){\mb(40,10){representation}}

\put(10,40){\line(1,0){30}}
\put(10,30){\line(1,0){20}}
\put(10,20){\line(1,0){20}}
\put(10,10){\line(1,0){30}}

\put(30,26){\fb(8,8){$H$}}
\put(30,16){\fb(8,8){$H$}}

\put(38,30){\line(1,0){2}}
\put(38,20){\line(1,0){2}}

\put(40,26){\framebox(10,8){$/$}}
\qbezier(40,28)(45,32)(50,28)
\put(50,30.5){\line(1,0){5}}
\put(50,29.5){\line(1,0){5}}
\put(55,26){\mb(10,8){$d_1$}}

\put(40,16){\framebox(10,8){$/$}}
\qbezier(40,18)(45,22)(50,18)
\put(50,20.5){\line(1,0){5}}
\put(50,19.5){\line(1,0){5}}
\put(55,16){\mb(10,8){$d_2$}}

\put(0,25){\mb(10,10){$|+\>$}}
\put(0,15){\mb(10,10){$|+\>$}}

\put(12,40){\mb(6,6){$1$}}
\put(12,30){\mb(6,6){$2$}}
\put(12,20){\mb(6,6){$3$}}
\put(12,10){\mb(6,6){$4$}}

\put(25,30){\line(0,1){10}}

\put(23,28){\mb(4,4){$\times$}}
\put(23,38){\mb(4,4){$\times$}}

\put(20,20){\line(0,1){10}}

\put(18,18){\mb(4,4){$\times$}}
\put(18,28){\mb(4,4){$\times$}}

\put(25,10){\line(0,1){10}}

\put(23,8){\mb(4,4){$\times$}}
\put(23,18){\mb(4,4){$\times$}}

\ep     
\label{eq:rcz}
\eea
Note that we have explicitly labeled all the qubits.  
The circuit in \eq{rcz} starts with a graph state, and applies the
gate $Z^{d_2} \ot Z^{d_1} \Lambda(Z)$ to qubits 1 and 4.
On the other hand, the deletion principle from \cite{Raussendorf01a}
(see \sec{1wqcmodel}) shows that, if the $H$ gates on qubits 2 and 3 are simply
omitted, and those qubits are measured along $\{|0\>, |1\>\}$, then qubits 1
and 4 are disentangled, and an identity gate is simulated instead.

Substituting the initial graph state of~\eq{rcz} for the undoable
$\Lambda(Z)$ in \eq{1wqctg}, the initial graph state for our first
1WQC$_{\rm T}$ scheme is given by
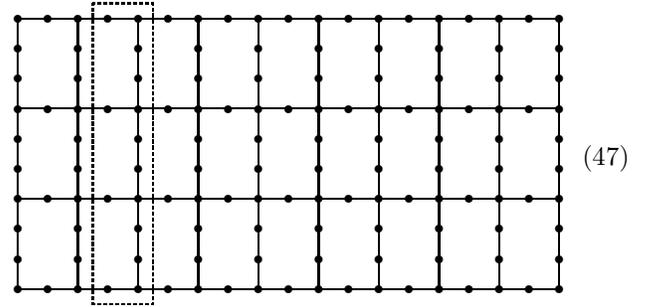
\bea
\setlength{\unitlength}{0.4mm}
\centering
\bp(180,100)
\multiput(0,5)(0,30){4}{\line(1,0){180}}
\multiput(0,5)(20,00){10}{\line(0,1){90}}
\multiput(0,5)(0,30){4}{   \multiput(0,0)(10,00){19}{\circle*{3}}  }
\multiput(0,15)(0,30){3}{  \multiput(0,0)(20,00){10}{\circle*{3}} 
                           \multiput(0,10)(20,00){10}{\circle*{3}}  }
\put(25,0){\dashbox(20,100){}}
\ep
\label{eq:remoteczapproach} 
\eea
This graph state can be used to simulate a $4$-qubit circuit for $9$ cycles
of ($i$) arbitrary $X$ rotations and ($ii$) arbitrary $Z$ rotations
and optional nearest-neighbor
$\Lambda(Z)$.  The region corresponding to the simulation of one cycle 
is marked by a dashed box.
The above state can in turn be obtained by ``deleting'' qubits denoted
by empty circles (by measuring them in the computational basis) in the
following cluster state:
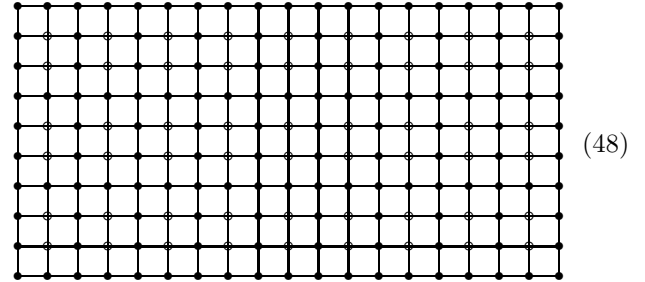
\bea
\setlength{\unitlength}{0.4mm}
\centering
\bp(180,90)
\multiput(0,0)(0,10){10}{\line(1,0){180}}
\multiput(0,0)(10,00){19}{\line(0,1){90}}
\multiput(0,0)(0,30){4}{  \multiput(0,0)(10,00){19}{\circle*{3}}  }
\multiput(0,10)(0,30){3}{ \multiput(0,0)(20,00){10}{\circle*{3}} 
                          \multiput(0,10)(20,00){10}{\circle*{3}} }
\multiput(10,10)(0,30){3}{  \multiput(0,0)(20,00){9}{\circle{3}}
                            \multiput(0,10)(20,00){9}{\circle{3}} }
\ep
\eea
This cluster state-based simulation requires six physical qubits
per logical qubit per cycle.

%%%%%%%%%%%%%%%%%%%%%%%%%%%%%%%%%%%%%%%%%%%%%%%%%%%%%%%%%%%%%%%%%%%%%%%%%
\subsubsection{The remote $\Lambda(Z)$ approach (II)}

The remote $\Lambda(Z)$ described in the previous section requires two
ancilla qubits per $\Lambda(Z)$.
We can save resources by relaxing the exact simulation condition, 
and use the following circuit:
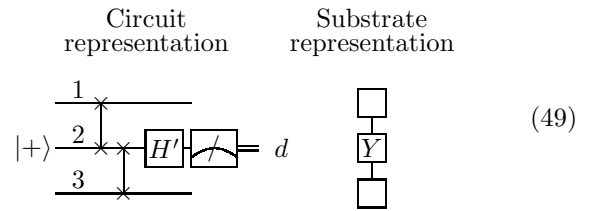
\bea
\setlength{\unitlength}{0.6mm}
\centering
\bp(100,55)

\put(80,23){\line(0,1){4}}
\put(80,13){\line(0,1){4}}

\put(77,27){\fb(6,6){}}
\put(77,17){\fb(6,6){$Y$}}
\put(77, 7){\fb(6,6){}}

\put(0,44){\mb(60,10){Circuit}}
\put(0,38){\mb(60,10){representation}}
\put(60,44){\mb(40,10){Substrate}}
\put(60,38){\mb(40,10){representation}}

\put(10,30){\line(1,0){30}}
\put(10,20){\line(1,0){20}}
\put(10,10){\line(1,0){30}}

\put(30,16){\fb(8,8){$H'$}}

\put(38,20){\line(1,0){2}}

\put(40,16){\framebox(10,8){$/$}}
\qbezier(40,18)(45,22)(50,18)
\put(50,20.5){\line(1,0){5}}
\put(50,19.5){\line(1,0){5}}
\put(55,16){\mb(10,8){$d$}}

\put(0,15){\mb(10,10){$|+\>$}}

\put(12,30){\mb(6,6){$1$}}
\put(12,20){\mb(6,6){$2$}}
\put(12,10){\mb(6,6){$3$}}

\put(20,20){\line(0,1){10}}

\put(18,18){\mb(4,4){$\times$}}
\put(18,28){\mb(4,4){$\times$}}

\put(25,10){\line(0,1){10}}

\put(23,8){\mb(4,4){$\times$}}
\put(23,18){\mb(4,4){$\times$}}

\ep     
\label{eq:rcz2}
\eea
where $H'= \smfrac{1}{\sqrt{2}} (Z+Y)$ and $H'$ followed by $M_Z$ is
simply $M_Y$.  Let $R_d$ be the $Z$-rotation $Z_{(-1)^{d+1}
\pi/2}$.  The above circuit implements the gate $(R_d \ot
R_d) \Lambda(Z)$, and yields a 1WQC$_{T}$ scheme from
the 1WQC$_{TG}$ model in \sec{1bit-telep}, because arbitrary $Z$
rotations are always simulated with the undoable $\Lambda(Z)$.  
Substituting \eq{rcz2} into \eq{1wqctg} gives another universal
initial graph state
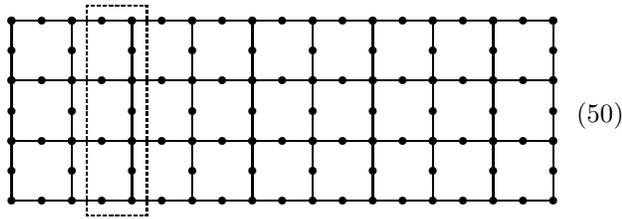
\bea
\setlength{\unitlength}{0.4mm}
\centering
\bp(180,70)
\multiput(0,5)(0,20){4}{\line(1,0){180}}
\multiput(0,5)(20,00){10}{\line(0,1){60}}
\multiput(0,5)(0,20){4}{   \multiput(0,0)(10,00){19}{\circle*{3}}  }
\multiput(0,15)(0,20){3}{  \multiput(0,0)(20,00){10}{\circle*{3}} 
                           \multiput(0,10)(20,00){10}{\circle*{3}}  }
\put(25,0){\dashbox(20,70){}}
\ep
\eea
which can again be obtained from the cluster state by deleting the cites
marked by empty circles:
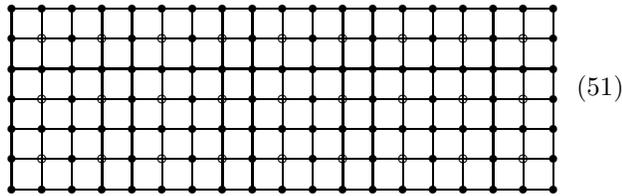
\bea
\setlength{\unitlength}{0.4mm}
\centering
\bp(180,70)
\multiput(0,0)(0,10){7}{\line(1,0){180}}
\multiput(0,0)(10,00){19}{\line(0,1){60}}
\multiput(0,0)(0,20){4}{  \multiput(0,0)(10,00){19}{\circle*{3}}  }
\multiput(0,10)(0,20){3}{ \multiput(0,0)(20,00){10}{\circle*{3}} }
\multiput(10,10)(0,20){3}{  \multiput(0,0)(20,00){9}{\circle{3}}}
\ep
\eea
This scheme requires only four physical qubits per logical qubit per
cycle.  

The above schemes are quite efficient in terms of the number of physical
qubits per logical operation.  Moreover, such efficiency is achieved with
remarkably little manipulation; rather, it arises from following simple
guidelines in a systematic derivation.

%%%%%%%%%%%%%%%%%%%%%%%%%%%%%%%%%%%%%%%%%%%%%%%%%%%%%%%%%%%%%%%%%%%%%%%%%
\subsubsection{The cancellation approach}

The cancellation approach uses the fact that in 1WQC$_{\rm TG}$, the
angle of single-qubit rotations can be entirely determined by the
measurement basis, and can be chosen on-line.
The idea is to intersperse two consecutive $\Lambda(Z)$ gates with a
single-qubit rotation so that proper choices of the angle of rotation
allow the two interactions to add up or cancel out.  In particular, 
\bea
        \Lambda(Z) (I \! \ot \! X_\t) \Lambda(Z) = \left\{
        \begin{array}{ll} 
        ~~I & {\rm if~} \t = 0 \\
        \lpm I \! \ot \! X_{\pi/2} \rpm 
        \Lambda(X) & {\rm if~} \t = -\smfrac{\pi}{2} 
        \end{array} 
        \right.
\non
\eea
The gate $\Lambda = \lpm I \! \ot \! X_{\pi/2} \rpm \Lambda(X)$ is locally
equivalent to $\Lambda(X)$, and is universal given the ability to perform
all single-qubit gates.  
Thus, we can perform undoable nearest-neighbor $\Lambda$ gates by
introducing pairs of $\Lambda(Z)$ with a variable $X$ rotation interspersed
between each pair.  This observation can be used to extend 1WQC$_{\rm
TG}$ to a scheme with a fixed initial state.  The idea is to represent
a quantum computation as a set of repeating subunits.  A subunit
consists of the following: 
\begin{enumerate}
\setlength{\itemsep}{-.5ex}
\renewcommand{\labelenumi}{($\roman{enumi}$)}
\item arbitrary $X$ rotations, 
\item arbitrary $Z$ rotations and $\bigotimes_{i {\rm \,odd}} \Lambda(Z)_{i,i{+}1}$,
\item interspersed $X$ rotations, 
\item arbitrary $Z$ rotations and $\bigotimes_{i {\rm \,odd}} \Lambda(Z)_{i,i{+}1}$,
\item arbitrary $X$ rotations, 
\item arbitrary $Z$ rotations and $\bigotimes_{i{\rm \,even}} \Lambda(Z)_{i,i{+}1}$,
\item interspersed $X$ rotations, and 
\item arbitrary $Z$ rotations and $\bigotimes_{i{\rm \,even}} \Lambda(Z)_{i,i{+}1}$,
\end{enumerate}
where subscripts on $\Lambda(Z)$ denote the qubits being acted on.
Following the discussion in \sec{1bit-telep}, the initial graph state
is given by
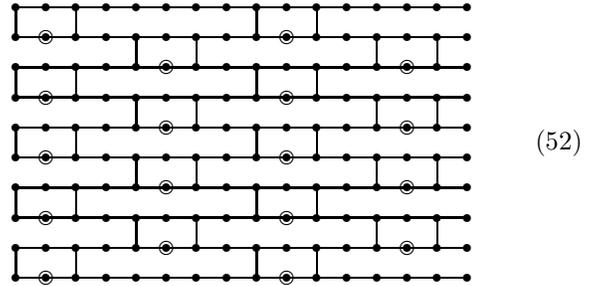
\bea
\setlength{\unitlength}{0.4mm}
\centering
\bp(150,93)
\multiput(0,0)(0,10){10}{\line(1,0){150}}
\multiput(0,0)(0,10){10}{  \multiput(0,0)(10,00){16}{\circle*{3}}  }

\multiput(10,0)(80,0){2}{
 \multiput(00,0)(0,20){5}{\circle{4}}
}

\multiput(50,10)(80,0){2}{
 \multiput(00,0)(0,20){4}{\circle{4}}
}

\multiput(00,0)(80,0){2}{
 \multiput(00,0)(0,20){5}{
  \multiput(0,0)(20,00){2}{\line(0,1){10}}
 }
}

\multiput(40,10)(80,0){2}{
 \multiput(00,0)(0,20){4}{
  \multiput(0,0)(20,00){2}{\line(0,1){10}}
 }
}
\ep
\eea
In this diagram, qubits corresponding to the interspersed $X$ rotations
are marked by extra circles.  The above state can be used to simulate
a five-qubit circuit for two cycles. 

The above graph state can also be produced from a cluster state,
though the resulting scheme is not as efficient as the first approach.

%%%%%%%%%%%%%%%%%%%%%%%%%%%%%%%%%%%%%%%%%%%%%%%%%%%%%%%%%%%%%%%%%%%%%%%%%
\subsubsection{The routing approach}

In the routing approach, a qubit can be teleported to an interaction
or a non-interaction site.  The interaction is always applied at the
interaction site, but it may be applied to the quantum data or to some
dummy state.

Using $X$ teleportation, it is possible to teleport a given input
state to one of several possible destinations.  To see how this works,
we consider the case of two destinations:
\bea
\setlength{\unitlength}{0.6mm}
\bp(120,32)
\put(-2,20){\makebox(12,10){$|\psi\rangle$}}
\put(10,25){\line(1,0){35}}

\put(0,10){\makebox(10,10){$|0\>$}}

\put(10,15){\line(1,0){5}}

\put(30,28){\line(0,-1){13}}
\put(30,25){\circle{6}}

\put(15,11){\fb(8,8){$H$}}
\put(23,15){\line(1,0){22}}

% ---------------------------------

\put(0,0){\makebox(10,10){$|0\>$}}

\put(10,5){\line(1,0){5}}

\put(37,28){\line(0,-1){23}}
\put(37,25){\circle{6}}

\put(15,1){\fb(8,8){$H$}}
\put(23,5){\line(1,0){17}}

\put(40,1){\framebox(12,8){$/$}}
\qbezier(40,3)(46,10)(52,3)
\put(52,4.5){\line(1,0){5}}
\put(52,5.5){\line(1,0){5}}
\put(57,0){\mb(5,10){$k$}}

\put(60,10){\mb(5,10){$=$}}

% -----------------------------------------------
\put(68,20){\makebox(12,10){$|\psi\rangle$}}
\put(80,25){\line(1,0){25}}
\put(105,21){\fb(8,8){$X^k$}}

\put(70,10){\makebox(10,10){$|0\>$}}

\put(80,15){\line(1,0){5}}

\put(113,25){\line(1,0){3}}

\put(97,28){\line(0,-1){13}}
\put(97,25){\circle{6}}

\put(85,11){\fb(8,8){$H$}}
\put(93,15){\line(1,0){23}}

\ep
\label{eq:xt-routing} 
\eea
Examining this circuit identity and comparing with the circuit for
$X$-teleportation, we see that by measuring the third qubit we can effect
an $X$-teleportation of the first qubit to the second.  Alternately, if
we had decided instead to measure the second qubit, we would have been
able to effect an $X$-teleportation of the first qubit to the third.
Thus, we are able to choose to route the state $|\psi\rangle$ to one
of two destinations.  The other qubit will be in a known state
$|k\rangle = |0\rangle$ or $|1\rangle$.
After this $X$-teleportation, the next simulation step is a
$Z$-teleportation that will perform $\Lambda(Z)$ on the path meant for
interaction and $I$ on the other path.  $Z$ rotations are also
performed at the same time.  In the previous step, the qubit state was
teleported to the desired destination, and the unwanted destination is
in some known random state $|k\>$.  The $Z$-teleportation can also be
constructed to take its input from either location, using the
identity
\bea
\setlength{\unitlength}{0.6mm}
\bp(120,30)
\put(0,20){\makebox(10,10){$|k \rangle$}}
\put(0,10){\makebox(10,10){$|\psi\rangle$}}
\put(0,0){\makebox(10,10){$|0\>$}}

\put(10,25){\line(1,0){31}}
\put(10,15){\line(1,0){20}}
\put(10,5){\line(1,0){31}}

\put(25,15){\line(0,-1){13}}

\put(25,5){\circle{6}}

\put(18,25){\line(0,-1){23}}
\put(18,5){\circle{6}}

\put(30,11){\fb(8,8){$H$}}
\put(38,15){\line(1,0){3}}

\put(54,10){\mb(6,10){$=$}}

%---------------------------------------------
\put(68,10){\makebox(12,10){$|\psi\rangle$}}
\put(70,0){\makebox(10,10){$|0\>$}}

\put(80,15){\line(1,0){20}}
\put(80,5){\line(1,0){20}}
\put(100,1){\fb(8,8){$X^k$}}
\put(108,5){\line(1,0){7}}

\put(90,15){\line(0,-1){13}}
\put(90,5){\circle{6}}

\put(100,11){\fb(8,8){$H$}}
\put(108,15){\line(1,0){7}}

\ep
\label{eq:zt-routing} 
\eea
We can combine the teleportation steps as in \eq{1bitteleport-chain},
and we obtain a simplified circuit analogous to \eq{1bit-chain}.
The following graph state is the initial state for this routing approach: 
\bea
\setlength{\unitlength}{0.4mm}
\centering
\bp(170,125)
\multiput(0,0)(0,25){5}{
 \multiput(10,10)(0,20){1}{  \multiput(0,0)(20,00){9}{\circle*{3}}  }
 \multiput(20,0)(0,20){2}{  \multiput(0,0)(20,00){8}{\circle*{3}}  }
 \multiput(10,10)(20,0){8}{
  \multiput(0,0)(20,00){1}{\line(1,1){10}}
  \multiput(0,0)(20,00){1}{\line(1,-1){10}}
  \multiput(20,0)(20,00){1}{\line(-1,1){10}}
  \multiput(20,0)(20,00){1}{\line(-1,-1){10}}
 }
}

\thicklines

\multiput(20,20)(0,50){2}{
\multiput(0,0)(40,0){4}{\line(0,1){5}}   }
\multiput(40,45)(0,50){2}{
\multiput(0,0)(40,0){4}{\line(0,1){5}}   }
\ep
\label{eq:routingstate}
\eea
The processing of information in this graph state is easy to
understand.  Consider the top three lines of qubits in the graph,
i.e., the eight adjacent ``diamonds'' at the top of the graph.  This
line of diamonds represents the processing of a single logical qubit.
Information starts out in the leftmost vertex of the diamond, and is
then routed either to the top vertex of the diamond, or to the bottom
vertex.  If it is routed to the bottom vertex, then it may be
interacted with the second row of diamonds, representing the second
logical qubit, effecting a $\Lambda(Z)$ gate between logical qubits.
If it is routed through the top vertex, then no interaction takes
place.  Finally, Z teleportation is used to reroute the information
from either the top or the bottom vertex into the rightmost vertex of
the diamond.  Thus, we see that this state can be used to simulate a
five-qubit circuit for four cycles.

%%%%%%%%%%%%%%%%%%%%%%%%%%%%%%%%%%%%%%%%%%%%%%%%%%%%%%%%%%%%%%%%%%%%%%%%%
\section{Conclusion}
\label{sec:conclusion}

In this paper we have explained how one-bit teleportation can be used
as a simple underlying principle to systematically derive
measurement-based schemes for universal quantum computation.  These
derivations provide a single unified approach that encompasses schemes
similar to both the 1WQC (one-way quantum computer) model introduced
in~\cite{Raussendorf01a}, and the TQC (teleportation-based model of
quantum computation) introduced in~\cite{Nielsen01t}.  However, our
schemes have the added advantage of being significantly simpler than
previously known schemes in either approach.  Most importantly, our
derivation has elicited a simple underlying principle for the 1WQC.

We have also outlined a variety of tools and techniques for designing
schemes for measurement-based quantum computation.  Our schemes have
many variants, indicating the flexibility of our constructions.  We
hope that the library of tools we have described will be of use both
in developing further insight into the power and limitations of
measurement-based quantum computation, and in designing 1WQC schemes
suited to a particular information processing task or physical
implementation.

\bigskip

\acknowledgments

We thank Panos Aliferis for sharing many thoughts and unpublished
results.  We thank Michael Ben-Or, Hans Briegel, Isaac Chuang, Chris
Dawson, Steven van Enk, Christopher Fuchs, Daniel Gottesman, Ashwin
Nayak, Robert Raussendorf, and Frank Verstraete for stimulating
discussions.
We thank Hans Briegel and Robert Raussendorf for their permission to
use Figure~1.
We also thank Jozef Gruska for drawing our attention to the
independent result by Perdrix \cite{Perdrix04}.
AMC and MAN acknowledge the hospitality of the Caltech IQI, where this
work was initiated in August 2003.  Part of this work was done while
AMC and DWL were visiting the Perimeter Institute and while DWL was
visiting the University of Toronto in November 2003.  Their
hospitality is also much appreciated.

AMC received support from the Fannie and John Hertz Foundation, and
was also supported in part by the Cambridge--MIT Institute, by the
Department of Energy under cooperative research agreement
DE-FC02-94ER40818, and by the National Security Agency and Advanced
Research and Development Activity under Army Research Office contract
DAAD19-01-1-0656.
DWL received support from the Richard C.\ Tolman Endowment Fund, the
Croucher Foundation, and the US National Science Foundation under
grant no.~EIA-0086038. 
MAN received support from the Australian Research Council.

% \bibliographystyle{apsrev_title}
% \bibliographystyle{unsrt}
% \bibliography{meas}

\begin{thebibliography}{38}
\expandafter\ifx\csname natexlab\endcsname\relax\def\natexlab#1{#1}\fi
\expandafter\ifx\csname bibnamefont\endcsname\relax
  \def\bibnamefont#1{#1}\fi
\expandafter\ifx\csname bibfnamefont\endcsname\relax
  \def\bibfnamefont#1{#1}\fi
\expandafter\ifx\csname citenamefont\endcsname\relax
  \def\citenamefont#1{#1}\fi
\expandafter\ifx\csname url\endcsname\relax
  \def\url#1{\texttt{#1}}\fi
\expandafter\ifx\csname urlprefix\endcsname\relax\def\urlprefix{URL }\fi
\providecommand{\bibinfo}[2]{#2}
\providecommand{\eprint}[2][]{\url{#2}}

\bibitem[{\citenamefont{DiVincenzo}(1995)}]{DiVincenzo95a}
\bibinfo{author}{\bibfnamefont{D.~P.} \bibnamefont{DiVincenzo}},
  \emph{\bibinfo{title}{Quantum computation}}, \bibinfo{journal}{Science}
  \textbf{\bibinfo{volume}{270}}, \bibinfo{pages}{255} (\bibinfo{year}{1995}),
  \eprint{eprint {arXiv}:quant-ph/9503016}.

\bibitem[{\citenamefont{Preskill}(1998)}]{Preskill98bk}
\bibinfo{author}{\bibfnamefont{J.}~\bibnamefont{Preskill}},
  \emph{\bibinfo{title}{Physics 229, Advanced mathematical methods of physics:
  Quantum computation and information}} (\bibinfo{publisher}{Caltech},
  \bibinfo{address}{Pasadena, CA}, \bibinfo{year}{1998}),
  \urlprefix\url{http://www.theory.caltech.edu/people/preskill/ph229}.

\bibitem[{\citenamefont{Nielsen and Chuang}(2000)}]{Nielsen00bk}
\bibinfo{author}{\bibfnamefont{M.~A.} \bibnamefont{Nielsen}} \bibnamefont{and}
  \bibinfo{author}{\bibfnamefont{I.~L.} \bibnamefont{Chuang}},
  \emph{\bibinfo{title}{Quantum computation and quantum information}}
  (\bibinfo{publisher}{Cambridge University Press},
  \bibinfo{address}{Cambridge, U.K.}, \bibinfo{year}{2000}).

\bibitem[{\citenamefont{Bennett et~al.}(1993)\citenamefont{Bennett, Brassard,
  Cr$\acute{e}$peau, Jozsa, Peres, and Wootters}}]{Bennett93}
\bibinfo{author}{\bibfnamefont{C.~H.} \bibnamefont{Bennett}},
  \bibinfo{author}{\bibfnamefont{G.}~\bibnamefont{Brassard}},
  \bibinfo{author}{\bibfnamefont{C.}~\bibnamefont{Cr$\acute{e}$peau}},
  \bibinfo{author}{\bibfnamefont{R.}~\bibnamefont{Jozsa}},
  \bibinfo{author}{\bibfnamefont{A.}~\bibnamefont{Peres}}, \bibnamefont{and}
  \bibinfo{author}{\bibfnamefont{W.}~\bibnamefont{Wootters}},
  \emph{\bibinfo{title}{Teleporting an unknown quantum state via dual classical
  and \mbox{Einstein}-\mbox{Podolsky}-\mbox{Rosen} channels}},
  \bibinfo{journal}{Phys. Rev. Lett.} \textbf{\bibinfo{volume}{70}},
  \bibinfo{pages}{1895} (\bibinfo{year}{1993}).

\bibitem[{\citenamefont{Shor}(1995)}]{Sho95}
\bibinfo{author}{\bibfnamefont{P.}~\bibnamefont{Shor}},
  \emph{\bibinfo{title}{Scheme for reducing decoherence in quantum computer
  memory}}, \bibinfo{journal}{Phys. Rev. A} \textbf{\bibinfo{volume}{52}},
  \bibinfo{pages}{2493} (\bibinfo{year}{1995}).

\bibitem[{\citenamefont{Shor}(1996)}]{Shor96}
\bibinfo{author}{\bibfnamefont{P.}~\bibnamefont{Shor}},
  \emph{\bibinfo{title}{Fault-tolerant quantum computation}}, in
  \emph{\bibinfo{booktitle}{Proc. 37$^{th}$ Annual Symposium on Foundations of
  Computer Science}} (\bibinfo{publisher}{IEEE Computer Society Press},
  \bibinfo{address}{Los Alamitos, CA}, \bibinfo{year}{1996}),
  p.~\bibinfo{pages}{56}, \eprint{eprint {arXiv}:quant-ph/9605011}.

\bibitem[{\citenamefont{Boykin et~al.}(1999)\citenamefont{Boykin, Mor, Pulver,
  Roychowdhury, and Vatan}}]{Boykin99}
\bibinfo{author}{\bibfnamefont{P.~O.} \bibnamefont{Boykin}},
  \bibinfo{author}{\bibfnamefont{T.}~\bibnamefont{Mor}},
  \bibinfo{author}{\bibfnamefont{M.}~\bibnamefont{Pulver}},
  \bibinfo{author}{\bibfnamefont{V.}~\bibnamefont{Roychowdhury}},
  \bibnamefont{and} \bibinfo{author}{\bibfnamefont{F.}~\bibnamefont{Vatan}},
  \emph{\bibinfo{title}{On universal and fault-tolerant quantum computing}}, in
  \emph{\bibinfo{booktitle}{Proc. 40$^{th}$ Annual Symposium on Foundations of
  Computer Science}} (\bibinfo{publisher}{IEEE Computer Society Press},
  \bibinfo{address}{Los Alamitos, CA}, \bibinfo{year}{1999}), \eprint{eprint
  {arXiv}:quant-ph/9906054}.

\bibitem[{\citenamefont{Knill et~al.}(1998)\citenamefont{Knill, Laflamme, and
  Zurek}}]{Knill98}
\bibinfo{author}{\bibfnamefont{E.}~\bibnamefont{Knill}},
  \bibinfo{author}{\bibfnamefont{R.}~\bibnamefont{Laflamme}}, \bibnamefont{and}
  \bibinfo{author}{\bibfnamefont{W.}~\bibnamefont{Zurek}},
  \emph{\bibinfo{title}{Resilient quantum computation}},
  \bibinfo{journal}{Science} \textbf{\bibinfo{volume}{279}},
  \bibinfo{pages}{342} (\bibinfo{year}{1998}), \eprint{eprint
  {arXiv}:quant-ph/9702058}.

\bibitem[{\citenamefont{Gottesman and Chuang}(1999)}]{Gottesman99t}
\bibinfo{author}{\bibfnamefont{D.}~\bibnamefont{Gottesman}} \bibnamefont{and}
  \bibinfo{author}{\bibfnamefont{I.~L.} \bibnamefont{Chuang}},
  \emph{\bibinfo{title}{Demonstrating the viability of universal quantum
  computation using teleportation and single-qubit operations}},
  \bibinfo{journal}{Nature} \textbf{\bibinfo{volume}{402}},
  \bibinfo{pages}{390} (\bibinfo{year}{1999}), \eprint{eprint
  {arXiv}:quant-ph/9908010}.

\bibitem[{\citenamefont{Zhou et~al.}(2000)\citenamefont{Zhou, Leung, and
  Chuang}}]{Zhou00}
\bibinfo{author}{\bibfnamefont{X.}~\bibnamefont{Zhou}},
  \bibinfo{author}{\bibfnamefont{D.}~\bibnamefont{Leung}}, \bibnamefont{and}
  \bibinfo{author}{\bibfnamefont{I.}~\bibnamefont{Chuang}},
  \emph{\bibinfo{title}{Methodology for quantum logic gate construction}},
  \bibinfo{journal}{Phys. Rev. A} \textbf{\bibinfo{volume}{62}},
  \bibinfo{pages}{052316} (\bibinfo{year}{2000}), \eprint{eprint
  {arXiv}:quant-ph/0002039}.

\bibitem[{\citenamefont{Knill et~al.}(2001)\citenamefont{Knill, Laflamme, and
  Milburn}}]{Knill01a}
\bibinfo{author}{\bibfnamefont{E.}~\bibnamefont{Knill}},
  \bibinfo{author}{\bibfnamefont{R.}~\bibnamefont{Laflamme}}, \bibnamefont{and}
  \bibinfo{author}{\bibfnamefont{G.}~\bibnamefont{Milburn}},
  \emph{\bibinfo{title}{Efficient linear optics quantum computation}},
  \bibinfo{journal}{Nature} \textbf{\bibinfo{volume}{409}}, \bibinfo{pages}{46}
  (\bibinfo{year}{2001}), \eprint{eprint {arXiv}:quant-ph/0006088}.

\bibitem[{\citenamefont{Wu and Lidar}(2003)}]{Lidar03}
\bibinfo{author}{\bibfnamefont{L.-A.} \bibnamefont{Wu}} \bibnamefont{and}
  \bibinfo{author}{\bibfnamefont{D.~A.} \bibnamefont{Lidar}},
  \emph{\bibinfo{title}{Universal quantum computation using exchange
  interactions and teleportation of single-qubit operations}},
  \bibinfo{journal}{Phys. Rev. A} \textbf{\bibinfo{volume}{67}},
  \bibinfo{pages}{050303} (\bibinfo{year}{2003}), \eprint{eprint
  {arXiv}:quant-ph/0208118}.

\bibitem[{\citenamefont{Raussendorf and Briegel}(2001)}]{Raussendorf01a}
\bibinfo{author}{\bibfnamefont{R.}~\bibnamefont{Raussendorf}} \bibnamefont{and}
  \bibinfo{author}{\bibfnamefont{H.~J.} \bibnamefont{Briegel}},
  \emph{\bibinfo{title}{A one-way quantum computer}}, \bibinfo{journal}{Phys.
  Rev. Lett.} \textbf{\bibinfo{volume}{86}}, \bibinfo{pages}{5188}
  (\bibinfo{year}{2001}), \eprint{eprint {arXiv}:quant-ph/0010033}.

\bibitem[{\citenamefont{Briegel and Raussendorf}(2001)}]{Raussendorf00}
\bibinfo{author}{\bibfnamefont{H.~J.} \bibnamefont{Briegel}} \bibnamefont{and}
  \bibinfo{author}{\bibfnamefont{R.}~\bibnamefont{Raussendorf}},
  \emph{\bibinfo{title}{Persistent entanglement in arrays of interacting
  particles}}, \bibinfo{journal}{Phys. Rev. Lett.}
  \textbf{\bibinfo{volume}{86}}, \bibinfo{pages}{910} (\bibinfo{year}{2001}),
  \eprint{eprint {arXiv}:quant-ph/0004051}.

\bibitem[{\citenamefont{Nielsen}(2003{\natexlab{a}})}]{Nielsen01t}
\bibinfo{author}{\bibfnamefont{M.~A.} \bibnamefont{Nielsen}},
  \emph{\bibinfo{title}{Universal quantum computation using only projective
  measurement, quantum memory, and preparation of the $0$ state}},
  \bibinfo{journal}{Phys. Lett. A} \textbf{\bibinfo{volume}{308}},
  \bibinfo{pages}{96} (\bibinfo{year}{2003}{\natexlab{a}}), \eprint{eprint
  {arXiv}:quant-ph/0108020}.

\bibitem[{\citenamefont{Nielsen and Chuang}(1997)}]{Nielsen97c}
\bibinfo{author}{\bibfnamefont{M.~A.} \bibnamefont{Nielsen}} \bibnamefont{and}
  \bibinfo{author}{\bibfnamefont{I.~L.} \bibnamefont{Chuang}},
  \emph{\bibinfo{title}{Programmable quantum gate arrays}},
  \bibinfo{journal}{Phys. Rev. Lett.} \textbf{\bibinfo{volume}{79}},
  \bibinfo{pages}{321} (\bibinfo{year}{1997}), \eprint{eprint
  {arXiv}:quant-ph/9703032}.

\bibitem[{\citenamefont{Fenner and Zhang}()}]{Fenner01}
\bibinfo{author}{\bibfnamefont{S.~A.} \bibnamefont{Fenner}} \bibnamefont{and}
  \bibinfo{author}{\bibfnamefont{Y.}~\bibnamefont{Zhang}},
  \emph{\bibinfo{title}{Universal quantum computation with two- and three-qubit
  projective measurements}}, \eprint{eprint {arXiv}:quant-ph/0111077}.

\bibitem[{\citenamefont{Leung}({\natexlab{a}})}]{Leung01c}
\bibinfo{author}{\bibfnamefont{D.~W.} \bibnamefont{Leung}},
  \emph{\bibinfo{title}{Two-qubit projective measurements are universal for
  quantum computation}}, \eprint{eprint {arXiv}:quant-ph/0111122}.

\bibitem[{\citenamefont{Leung}(2004)}]{Leung03t}
\bibinfo{author}{\bibfnamefont{D.~W.} \bibnamefont{Leung}},
  \emph{\bibinfo{title}{Quantum computation by measurements}},
  \bibinfo{journal}{Internation Journal of Quantum Information}
  \textbf{\bibinfo{volume}{2}}, \bibinfo{pages}{33} (\bibinfo{year}{2004}),
  \eprint{eprint {arXiv}:quant-ph/0310189}.

\bibitem[{\citenamefont{Raussendorf and Briegel}(2002)}]{Raussendorf02a}
\bibinfo{author}{\bibfnamefont{R.}~\bibnamefont{Raussendorf}} \bibnamefont{and}
  \bibinfo{author}{\bibfnamefont{H.~J.} \bibnamefont{Briegel}},
  \emph{\bibinfo{title}{Computational model underlying the one-way quantum
  computer}}, \bibinfo{journal}{Quantum Information and Computation}
  \textbf{\bibinfo{volume}{2}}, \bibinfo{pages}{443} (\bibinfo{year}{2002}),
  \eprint{eprint {arXiv}:quant-ph/0108067}.

\bibitem[{\citenamefont{Gottesman}(1996)}]{Gottesman96}
\bibinfo{author}{\bibfnamefont{D.}~\bibnamefont{Gottesman}},
  \emph{\bibinfo{title}{A class of quantum error-correcting codes saturating
  the quantum \mbox{Hamming} bound}}, \bibinfo{journal}{Phys. Rev. A}
  \textbf{\bibinfo{volume}{54}}, \bibinfo{pages}{1862} (\bibinfo{year}{1996}),
  \eprint{eprint {arXiv}:quant-ph/9604038}.

\bibitem[{\citenamefont{Nielsen}()}]{Nielsen04}
\bibinfo{author}{\bibfnamefont{M.~A.} \bibnamefont{Nielsen}},
  \emph{\bibinfo{title}{Optical quantum computation using cluster states}},
  \bibinfo{note}{to appear in Phys. Rev. Lett.}, \eprint{eprint
  {arXiv}:quant-ph/0402005}.

\bibitem[{\citenamefont{Childs}()}]{ChildsPItalk03}
\bibinfo{author}{\bibfnamefont{A.~M.} \bibnamefont{Childs}},
  \emph{\bibinfo{title}{Teleportation-based approaches to universal quantum
  computation with single-qubit measurement}}, \bibinfo{howpublished}{Seminar
  at the Perimeter Institute, November 2003},
  \urlprefix\url{http://www.qinfo.org/qc-by-measurement}.

\bibitem[{\citenamefont{Leung}({\natexlab{b}})}]{LeungIQItalk04}
\bibinfo{author}{\bibfnamefont{D.~W.} \bibnamefont{Leung}},
  \emph{\bibinfo{title}{Systematic derivation of universal quantum computation
  schemes with one-qubit measurements and universal initial state based on
  teleportation}}, \bibinfo{howpublished}{Group meeting talk at the Institute
  for Quantum Information, Caltech, Feburary 2004},
  \urlprefix\url{http://www.qinfo.org/qc-by-measurement}.

\bibitem[{\citenamefont{Leung}({\natexlab{c}})}]{LeungERATOtalk04}
\bibinfo{author}{\bibfnamefont{D.~W.} \bibnamefont{Leung}},
  \emph{\bibinfo{title}{Unifying and simplifying measurement-based quantum
  computation schemes}}, \bibinfo{howpublished}{Seminar at ERATO Tokyo office,
  March 2004}, \urlprefix\url{http://www.qinfo.org/qc-by-measurement}.

\bibitem[{\citenamefont{Verstraete and Cirac}()}]{Verstraete03}
\bibinfo{author}{\bibfnamefont{F.}~\bibnamefont{Verstraete}} \bibnamefont{and}
  \bibinfo{author}{\bibfnamefont{J.~I.} \bibnamefont{Cirac}},
  \emph{\bibinfo{title}{Valence bond solids for quantum computation}},
  \eprint{eprint {arXiv}:quant-ph/0311130}.

\bibitem[{\citenamefont{Aliferis and Leung}()}]{Aliferis04}
\bibinfo{author}{\bibfnamefont{P.}~\bibnamefont{Aliferis}} \bibnamefont{and}
  \bibinfo{author}{\bibfnamefont{D.~W.} \bibnamefont{Leung}},
  \emph{\bibinfo{title}{Computation by measurements: a unifying picture}},
  \eprint{eprint {arXiv}:quant-ph/0404082}.

\bibitem[{\citenamefont{Jorrand and Perdrix}()}]{JP04}
\bibinfo{author}{\bibfnamefont{P.}~\bibnamefont{Jorrand}} \bibnamefont{and}
  \bibinfo{author}{\bibfnamefont{S.}~\bibnamefont{Perdrix}},
  \emph{\bibinfo{title}{Unifying quantum computation with projective
  measurements only and one-way quantum computation}}, \eprint{eprint
  {arXiv}:quant-ph/0404125}.

\bibitem[{\citenamefont{Perdrix}()}]{Perdrix04}
\bibinfo{author}{\bibfnamefont{S.}~\bibnamefont{Perdrix}},
  \emph{\bibinfo{title}{State transfer instead of teleportation in
  measurement-based quantum computation}}, \eprint{eprint
  {arXiv}:quant-ph/0402204}.

\bibitem[{\citenamefont{Perdrix and Jorrand}()}]{PJ04b}
\bibinfo{author}{\bibfnamefont{S.}~\bibnamefont{Perdrix}} \bibnamefont{and}
  \bibinfo{author}{\bibfnamefont{P.}~\bibnamefont{Jorrand}},
  \emph{\bibinfo{title}{Measurement-based quantum turing machines and their
  universality}}, \eprint{eprint {arXiv}:quant-ph/0404146}.

\bibitem[{\citenamefont{Browne and Rudolph}()}]{Browne04}
\bibinfo{author}{\bibfnamefont{D.~E.} \bibnamefont{Browne}} \bibnamefont{and}
  \bibinfo{author}{\bibfnamefont{T.}~\bibnamefont{Rudolph}},
  \emph{\bibinfo{title}{Efficient linear optical quantum computation}},
  \eprint{eprint {arXiv}:quant-ph/0405157}.

\bibitem[{\citenamefont{Nielsen and Dawson}()}]{ND04}
\bibinfo{author}{\bibfnamefont{M.~A.} \bibnamefont{Nielsen}} \bibnamefont{and}
  \bibinfo{author}{\bibfnamefont{C.~M.} \bibnamefont{Dawson}},
  \emph{\bibinfo{title}{Fault-tolerant quantum computation with cluster
  states}}, \eprint{eprint {arXiv}:quant-ph/0405134}.

\bibitem[{\citenamefont{Raussendorf et~al.}(2003)\citenamefont{Raussendorf,
  Browne, and Briegel}}]{Raussendorf03}
\bibinfo{author}{\bibfnamefont{R.}~\bibnamefont{Raussendorf}},
  \bibinfo{author}{\bibfnamefont{D.~E.} \bibnamefont{Browne}},
  \bibnamefont{and} \bibinfo{author}{\bibfnamefont{H.~J.}
  \bibnamefont{Briegel}}, \emph{\bibinfo{title}{Measurement-based quantum
  computation with cluster states}}, \bibinfo{journal}{Phys. Rev. A}
  \textbf{\bibinfo{volume}{68}}, \bibinfo{pages}{022312}
  (\bibinfo{year}{2003}), \eprint{eprint {arXiv}:quant-ph/0301052}.

\bibitem[{\citenamefont{Schlingemann and Werner}(2001)}]{Werner01}
\bibinfo{author}{\bibfnamefont{D.}~\bibnamefont{Schlingemann}}
  \bibnamefont{and} \bibinfo{author}{\bibfnamefont{R.~F.}
  \bibnamefont{Werner}}, \emph{\bibinfo{title}{Quantum error-correcting codes
  associated with graphs}}, \bibinfo{journal}{Phys. Rev. A}
  \textbf{\bibinfo{volume}{65}}, \bibinfo{pages}{012308}
  (\bibinfo{year}{2001}), \eprint{eprint {arXiv}:quant-ph/0012111}.

\bibitem[{\citenamefont{Hein et~al.}(2004)\citenamefont{Hein, Eisert, and
  Briegel}}]{Briegel03}
\bibinfo{author}{\bibfnamefont{M.}~\bibnamefont{Hein}},
  \bibinfo{author}{\bibfnamefont{J.}~\bibnamefont{Eisert}}, \bibnamefont{and}
  \bibinfo{author}{\bibfnamefont{H.~J.} \bibnamefont{Briegel}},
  \emph{\bibinfo{title}{Multi-party entanglement in graph states}},
  \bibinfo{journal}{Phys. Rev. A} \textbf{\bibinfo{volume}{69}},
  \bibinfo{pages}{062311} (\bibinfo{year}{2004}), \eprint{eprint
  {arXiv}:quant-ph/0307130}.

\bibitem[{\citenamefont{Nielsen}(2003{\natexlab{b}})}]{Nielsen03rb}
\bibinfo{author}{\bibfnamefont{M.~A.} \bibnamefont{Nielsen}},
  \emph{\bibinfo{title}{Journal club notes on the cluster-state model of
  quantum computation}} (\bibinfo{year}{2003}{\natexlab{b}}),
  \urlprefix\url{http://www.qinfo.org/qc-by-measurement}.

\bibitem[{\citenamefont{Gottesman}(1999{\natexlab{a}})}]{Gottesman98h}
\bibinfo{author}{\bibfnamefont{D.}~\bibnamefont{Gottesman}},
  \emph{\bibinfo{title}{The \mbox{Heisenberg} representation of quantum
  computers}}, in \emph{\bibinfo{booktitle}{Proc. XXII International Colloquium
  on Group Theoretical Methods in Physics}}, edited by
  \bibinfo{editor}{\bibfnamefont{S.~P.} \bibnamefont{Corney}},
  \bibinfo{editor}{\bibfnamefont{R.}~\bibnamefont{Delbourgo}},
  \bibnamefont{and} \bibinfo{editor}{\bibfnamefont{P.~D.} \bibnamefont{Jarvis}}
  (\bibinfo{publisher}{International Press}, \bibinfo{address}{Cambridge, MA},
  \bibinfo{year}{1999}{\natexlab{a}}), pp. \bibinfo{pages}{32--33},
  \bibinfo{note}{eprint {arXiv}:quant-ph/9807006}.

\bibitem[{\citenamefont{Gottesman}(1999{\natexlab{b}})}]{Gottesman98d}
\bibinfo{author}{\bibfnamefont{D.}~\bibnamefont{Gottesman}},
  \emph{\bibinfo{title}{Fault-tolerant quantum computation with
  higher-dimensional systems}}, \bibinfo{journal}{Lect. Notes. Comp. Sci.}
  \textbf{\bibinfo{volume}{1509}}, \bibinfo{pages}{302}
  (\bibinfo{year}{1999}{\natexlab{b}}), \eprint{eprint
  {arXiv}:quant-ph/9802007}.

\end{thebibliography}

\end{document}